\documentclass[prl,final,amsmath,amssymb,nofootinbib,twocolumn]{revtex4-1}
\usepackage{epsfig}
\usepackage{bbm}
\usepackage{graphicx}
\usepackage{ragged2e}
\usepackage{datetime}

\usepackage{color}
\definecolor{a}{rgb}{    0.8941,     0,0.0980}
\definecolor{b}{rgb}{    0.9373,0.3294,0.0824}
\definecolor{c}{rgb}{         0,0.5059,0.2039}
\definecolor{d}{rgb}{    0.0784,0.0784,0.0784}
\definecolor{e}{rgb}{    0.1255,0.1216,0.4235}

\newif\iffigs
\figstrue

\newcommand{\ket}[1]{\ensuremath{| #1 \rangle}}
\newcommand{\bra}[1]{\ensuremath{\langle #1 |}}
\newcommand{\nn}{\ensuremath{\nonumber}}

\begin{document}
\title[An Open-System Quantum Simulator with Trapped Ions]{An Open-System Quantum Simulator with Trapped Ions}
\author{Julio T. Barreiro*$^{,1}$, Markus M\"uller*$^{,2,3}$,
 Philipp Schindler$^1$, Daniel Nigg$^1$, Thomas Monz$^1$, 
 Michael Chwalla$^{1,2}$, Markus Hennrich$^1$, Christian F. Roos$^{1,2}$, Peter
 Zoller$^{2,3}$ and Rainer Blatt$^{1,2}$} 
\affiliation{%
$^{1}$Institut f\"ur Experimentalphysik, Universit\"at Innsbruck,
  Technikerstr. 25, 6020 Innsbruck, Austria\\
$^{2}$Institut f\"ur Quantenoptik und Quanteninformation,
\"Osterreichische Akademie der Wissenschaften, Technikerstr. 21A, 6020 Innsbruck,
  Austria\\
$^{3}$Institut f\"ur Theoretische Physik,
  Universit\"at Innsbruck, Technikerstr. 25, 6020 Innsbruck,
  Austria\\
* These authors contributed equally to this work.}

\begin{abstract}
\textbf{The control of quantum systems is of fundamental scientific interest
  and promises powerful applications and technologies.  Impressive progress has
  been achieved in isolating the systems from the environment and coherently
  controlling their dynamics, as demonstrated by the creation and manipulation
  of entanglement in various physical systems.  However, for open quantum
  systems, engineering the dynamics of many particles by a controlled coupling
  to an environment remains largely unexplored. Here we report the first
  realization of a toolbox for simulating an open quantum system with up to five
  qubits.  Using a quantum computing architecture with trapped ions, we combine
  multi-qubit gates with optical pumping to implement coherent operations and
  dissipative processes.  We illustrate this engineering by the dissipative
  preparation of entangled states, the simulation of coherent many-body spin
  interactions and the quantum non-demolition measurement of multi-qubit
  observables.  By adding controlled dissipation to coherent operations, this
  work offers novel prospects for open-system quantum simulation and
  computation.}

\end{abstract}

\keywords{one two three}
\pacs{PACS number}

\volumeyear{year}
\volumenumber{number}
\issuenumber{number}
\eid{identifier}

\maketitle

Every quantum system is inevitably coupled to its surrounding environment.
Significant progress has been made in isolating systems from their enviroment
and coherently controlling the dynamics of several
qubits~\cite{ladd-nature-464-45,kimble-nature-453-1023,schoelkopf-nature-451-664,neeley-nature-467-570}.
These achievements have enabled the realization of high-fidelity quantum gates,
the implementation of small-scale quantum computing and communication devices
as well as the measurement-based probabilistic preparation of entangled states,
in atomic~\cite{saffman-rmp-82-2313,bloch-rmp-80-885},
photonic~\cite{obrien-science-318-1567} and solid-state
setups~\cite{clarke-nature-453-1031,hanson-rmp-79-1217,wrachtrup-jpcm-18-S807}.
In particular, successful demonstrations of quantum
simulators~\cite{feynman-IJTPhys-21-467,lloyd-science-273-1073}, which allow
one to mimic and study the dynamics of complex quantum systems, have been
reported~\cite{buluta-science-326-108}.

In contrast, controlling the more general dynamics of open systems amounts to
engineering both the Hamiltonian time evolution of the system as well as the
coupling to the environment. Although open-system dynamics in a many-body or
multi-qubit system are typically associated with
decoherence~\cite{myatt-nature-403-269,deleglise-nature-455-510,barreiro-natphys-aop},
the ability to design dissipation can be a useful resource.  For example,
controlled dissipation allows the preparation of a desired entangled state from
an arbitrary
state~\cite{krauter-arxiv-1006.4344,diehl-natphys-4-878,cho-arxiv-1008.4088} or
an enhanced sensitivity for precision
measurements~\cite{goldstein-arxiv-1001.0089}.  In a broader context, by
combining suitably chosen coherent and dissipative time steps, one can realize
the most general non-unitary open-system evolution of a many-particle system.
This engineering of the system-environment coupling generalizes the concept of
Hamiltonian quantum simulation to open quantum systems.  In addition, this
engineering enables the dissipative preparation and manipulation of many-body
states and quantum phases \cite{diehl-arXiv:1007.3420}, and also quantum
computation based on dissipation~\cite{verstraete-nphys-5-633}.

Here we provide the first experimental demonstration of a complete toolbox,
through coherent and dissipative manipulations of a multi-qubit system, to
control the dynamics of open systems.  In a string of trapped ions, each ion
encoding a qubit, we subdivide the qubits into ``system'' and ``environment''.
The system-environment coupling is then engineered through the universal set of
quantum operations available in ion-trap quantum
computers~\cite{haffner-physrep-469-155,home-science-325-1227} and a
dissipative mechanism based on optical pumping.

We first illustrate this engineering by dissipatively preparing a Bell state in
a 2+1 ion system, such that an initially fully mixed state is pumped into a
given Bell state.  Similarly, with 4+1 ions, we also dissipatively prepare a
4-qubit GHZ-state, which can be regarded as a minimal instance of Kitaev's
toric code \cite{Kitaev-annalsphys-303-2}.  Besides the dissipative elements,
we show coherent $n$-body interactions by implementing the fundamental building
block for 4-spin interactions.  In addition, we demonstrate a readout of
$n$-particle observables in a non-destructive way with a quantum-nondemolition
(QND) measurement of a $4$-qubit stabilizer operator. Altogether, our work
demonstrates all essential coherent and dissipative elements for controlling
general open-system dynamics.

\section{Open-System Quantum Dynamics and Bell-State ``Cooling''}
\label{sec:TheorySection}

The dynamics of an open quantum system $\mathrm{S}$ coupled to an environment
$\mathrm{E}$ can be described by the unitary transformation $\rho_{SE}\mapsto
U\rho_{SE}U^{\dagger}$, with $\rho_{SE}$ the joint density matrix of the
composite system $\mathrm{S+E}$. Thus, the reduced density operator of the
system will evolve as $\rho_{S}=\mathrm{Tr}_{E}U\rho_{SE}U^{\dagger}$.  The
time evolution of the system can also be described by a completely positive
Kraus map
\begin{equation}
\label{eq:KrausMap}
\rho_{S}\mapsto\mathcal{E}(\rho_{S})=\sum_{k}E_{k}\rho_{S}E_{k}^{\dagger}
\end{equation}
with $E_{k}$ operation elements satisfying
$\sum_{k}E_{k}^{\dag}E_{k}=1$~\cite{nielsen-book}.  If the system is decoupled
from the environment, the general map (\ref{eq:KrausMap}) reduces to $\rho_S
\mapsto U_S \rho_S U_S^\dagger$, with $U_S$ the unitary time evolution operator
acting only on the system.

Control of both coherent and dissipative dynamics is then achieved by finding
corresponding sequences of maps (\ref{eq:KrausMap}) specified by sets of
operation elements $\{E_k\}$ and engineering these sequences in the laboratory.
In particular, for the example of dissipative quantum state preparation,
pumping to an entangled state $|\psi\rangle$ reduces to implementing
appropriate sequences of dissipative maps.  These maps are chosen to drive the
system to the desired target state irrespective of its initial state.  The
resulting dynamics have then the pure state $|\psi\rangle$ as the unique
attractor, $\rho_{S}\mapsto|\psi\rangle\langle\psi|$. In quantum optics and
atomic physics, the techniques of optical pumping and laser cooling are
successfully used for the dissipative preparation of quantum states, although
on a \textit{single-particle} level.  The engineering of dissipative maps for
the preparation of entangled states can be seen as a generalization of this
concept of pumping and cooling in driven dissipative systems to a
\textit{many-particle} context.  To be concrete, we focus on dissipative
preparation of stabilizer states, which represent a large family of entangled
states, including graph states and error-correcting
codes~\cite{steane-nature-399-6732}.

We start by outlining the concept of Kraus map engineering for the simplest
non-trivial example of ``cooling'' a system of two qubits into a Bell
state. The Hilbert space of two qubits is spanned by the four Bell states
defined as $\ket{\Phi^{\pm}} = \frac{1}{\sqrt{2}} (\ket{00} \pm \ket{11})$ and
$\ket{\Psi^{\pm}} = \frac{1}{\sqrt{2}} (\ket{01} \pm \ket{10})$. Here,
$\ket{0}$ and $\ket{1}$ denote the computational basis of each qubit, and we
use the short-hand notation $\ket{00} = \ket{0}_1 \ket{0}_2$, for example.
These maximally entangled states are stabilizer states: the Bell state
$\ket{\Phi^+}$, for instance, is said to be \textit{stabilized} by the two
stabilizer operators $Z_1Z_2$ and $X_1X_2$, where $X$ and $Z$ denote the usual
Pauli matrices, as it is the only two-qubit state being an eigenstate of
eigenvalue +1 of these two commuting observables,
i.e.~$Z_1Z_2\ket{\Phi^+}=\ket{\Phi^+}$ and $X_1X_2\ket{\Phi^+}=\ket{\Phi^+}$.
In fact, each of the four Bell states is uniquely determined as an eigenstate
with eigenvalues $\pm1$ with respect to $Z_1Z_2$ and $X_1X_2$. The key idea of
cooling is that we can achieve dissipative dynamics which pump the system into
a particular Bell state, for example
$\rho_{S}\mapsto|\Psi^-\rangle\langle\Psi^-|$, by constructing two dissipative
maps, under which the two qubits are irreversibly transfered from the +1 into
the -1 eigenspaces of $Z_1Z_2$ and $X_1X_2$.

The dissipative maps are engineered with the aid of an ancilla "environment"
qubit~\cite{lloyd-pra-65-010101,duer-pra-78-052325} and a quantum circuit of
coherent and dissipative operations. The form and decomposition of these maps
into basic operations are discussed in Box 1.  The cooling dynamics are
determined by the probability of pumping from the +1 into the -1 stabilizer
eigenspaces, which can be directly controlled by varying the parameters in the
employed gate operations. For pumping with unit probability ($p=1$), the two
qubits reach the target Bell state --- regardless of their initial state ---
after only one cooling cycle, i.e.,~by a single application of each of the two
maps.  In contrast, when the pumping probability is small ($p \ll 1$), the
process can be regarded as the infinitesimal limit of the general map
(\ref{eq:KrausMap}). In this case, the system dynamics under a repeated
application of the cooling cycle are described by a master
equation~\cite{wiseman-book}
\begin{align} 
\dot{\rho}_{S}  & =-i[H_S,\rho_{S}]\label{eq:masterequation}\\
& +\sum_{k}\left(  c_{k}\rho_{S}c_{k}^{\dag}-\frac{1}{2}c_{k}^{\dagger}%
c_{k}\rho_{S}-\rho_{S} \frac{1}{2}c_{k}^{\dagger}c_{k}\right).  \nonumber
\end{align}
Here, $H_S$ is a system Hamiltonian, and $c_{k}$ are Lindblad operators
reflecting the system-environment coupling. For the purely dissipative maps
discussed here, $H_S=0$. Quantum jumps from the +1 into the -1 eigenspace of
$Z_1Z_2$ and $X_1X_2$ are mediated by a set of \textit{two-qubit} Lindblad
operators (see box 1 for details); here the system reaches the target Bell
state asymptotically after many cooling cycles.

\begin{figure*}[ht]
\begin{center}
\textbf{Box 1: Engineering dissipative open-system dynamics} 
\fbox{
\begin{minipage}[l]{0.97\textwidth}
	\begingroup
	\parfillskip=0pt
	\begin{minipage}[p]{0.47\textwidth}
	\justifying Dissipative dynamics which cool two qubits from an
        arbitrary initial state into the Bell state $\ket{\Psi^-}$ are realized
        by two maps that generate pumping from the +1 into the -1 eigenspaces
        of the stabilizer operators $Z_1Z_2$ and $X_1X_2$:
        \begin{center}
        \vspace{-2mm}
        \includegraphics{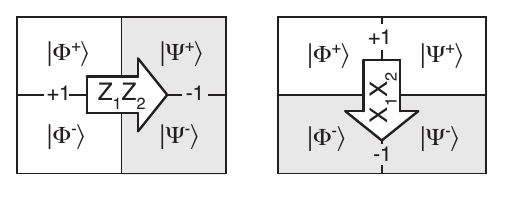}
        \end{center}
        \vspace{-4mm}
        \justifying \noindent For $Z_1Z_2$, the dissipative map pumping into the -1 eigenspace is
        $\rho_S \mapsto \mathcal{E}(\rho_S)=E_1 \rho_{S}E^\dagger_1 + E_2 \rho_{S}E^\dagger_2$ with 
	\begin{align}
	\label{eq:KrausOperators}
	E_{1} & = \sqrt{p} \, X_2 \frac{1}{2} \left(1 + Z_1Z_2 \right), \nn \\
	E_{2} & = \frac{1}{2} \left( 1 - Z_1Z_2 \right) +  \sqrt{1-p} \,\frac{1}{2} \left( 1 + Z_1Z_2 \right).\nonumber
	\end{align}
	\justifying The map's action as a uni-directional pumping process can
        be seen as follows.  Since the operation element $E_1$ contains the
        projector $\frac{1}{2} ( 1 + Z_1Z_2)$ onto the +1 eigenspace of
        $Z_1Z_2$, the spin flip $X_2$ can then convert +1 into -1 eigenstates
        of $Z_1Z_2$, e.g.,~$\ket{\Phi^+} \mapsto \ket{\Psi^+}$.  In contrast,
        the -1 eigenspace of $Z_1Z_2$ is left invariant.  In the limit $p\ll1$,
        the repeated application of this map reduces the process to a master
        equation with Lindblad operator $c=\frac{1}{2}X_2(1-Z_1Z_2)$.

	We implement the two dissipative maps by quantum circuits of three
        unitary operations (i)-(iii) and a dissipative step (iv).  Both maps
        act on the two system qubits $S$ and an ancilla which plays the role of
        the environment $E$:
	\begin{center}
        \vspace{-2mm}
	\includegraphics{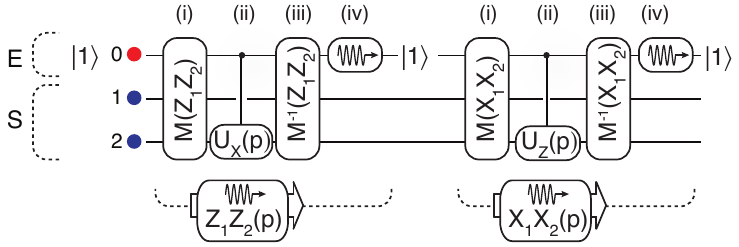}
        \end{center}

        \end{minipage}%
	\hfill
	\begin{minipage}[p]{0.47\textwidth}
	\justifying

Cooling $Z_1Z_2$ proceeds as follows:

        (i)~Information about whether the system is in the +1 or -1 eigenspace
of $Z_1Z_2$ is mapped by $M(Z_1Z_2)$ onto the logical states $\ket{0}$ and
$\ket{1}$ of the ancilla (initially in $\ket{1}$).

        (ii)~A controlled gate $C(p)$ converts +1 into -1 eigenstates by
        flipping the state of the second qubit with probability $p$, where
	\begin{equation}
	C(p) = \ket{0}\bra{0}_\mathrm{0} \otimes U_{X_2}(p) +
	\ket{1}\bra{1}_\mathrm{0} \otimes \openone,\nonumber
	\end{equation}
	with $U_{X_2}(p) = \exp (i \alpha X_2)$ and $p = \sin^2 \alpha$.

        (iii)~The initial mapping is inverted by $M^{-1}(Z_1Z_2)$.  At this
        stage, in general, the ancilla and system qubits are entangled.

        (iv) The ancilla is dissipatively reset to $\ket{1}$, which carries
        away entropy to ``cool'' the two system qubits.

      The second map for cooling into the -1 eigenspace of $X_1X_2$ is obtained
      from interchanging the roles of $X$ and $Z$ above.

	The engineering of dissipative maps can be readily generalized to
        systems of more qubits.  As an example, dissipative preparation of
        $n$-qubit stabilizer states can be realized by a sequence of $n$
        dissipative maps (e.g. for $Z_1Z_2$ and $X_1X_2X_3X_4$ pumping), which
        are implemented in analogy to the quantum circuits for Bell state
        cooling discussed above:
	\begin{center}
        \includegraphics{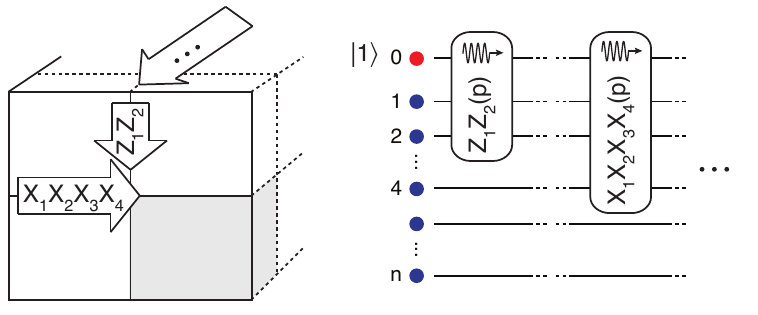}
        \end{center}
	\end{minipage}%
	\par
	\endgroup
\end{minipage}
}
\end{center}
\end{figure*}

\section{Experimental Bell-State Cooling}
\label{sec:ExpTools}

\begin{figure}[ht]
\begin{center}
\iffigs \includegraphics{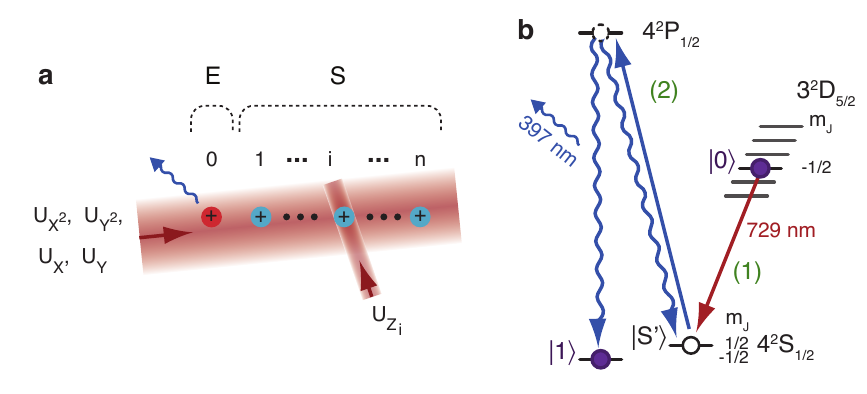}\fi
\end{center}
\caption{\textbf{Experimental tools for the simulation of open quantum systems
    with ions.}  \textbf{a,} The coherent component is realized by collective
  ($U_X,U_Y,U_{X^2},U_{Y^2}$) and single-qubit operations ($U_{Z_i}$) on a
  string of $^{40}$Ca$^+$ ions which consists of the environment qubit (ion 0)
  and the system qubits (ions 1 through n).  \textbf{b,} The dissipative
  mechanism on the ancilla qubit is realized in the two steps shown on the
  Zeeman-split $^{40}$Ca$^+$ levels by (1) a coherent transfer of the
  population from $|0\rangle$ to $|S'\rangle$ and (2) an optical pumping to
  $|1\rangle$ after a transfer to the $4^2P_{1/2}$ state by a
  circularly-polarised laser at 397 nm.}
\label{fig:FigureExperiment}
\end{figure}

The dissipative preparation of $n$-particle entangled states is realized in a
system of $n$+1 $^{40}$Ca$^+$ ions confined to a string by a linear Paul trap
and cooled to the ground state of the axial centre-of-mass
mode~\cite{schmidt-kaler-apblo-77-789}.  For each ion, the internal electronic
Zeeman levels $D_{5/2}(m=-1/2)$ and $S_{1/2}(m=-1/2)$ encode the logical states
$|0\rangle$ and $|1\rangle$ of a qubit.  For coherent operations, a laser at a
wavelength of 729~nm excites the quadrupole transition connecting the qubit
states ($S_{1/2}\leftrightarrow D_{5/2}$).  A broad beam of this laser couples
to all ions (see Fig.~\ref{fig:FigureExperiment}a) and realizes the collective
single-qubit gate $U_X(\theta)=\exp(-i\frac{\theta}{2}\sum_iX_i)$ as well as a
M\o{}lmer-S\o{}rensen~\cite{molmer-prl-82-1835} (MS) entangling operation
$U_{X^2}(\theta)=\exp(-i\frac{\theta}{4}(\sum_iX_i)^2)$ when using a
bichromatic light field~\cite{roos-njp-10-013002}.  Shifting the optical phase
of the drive field by $\pi/2$ exchanges $X_i$ by $Y_i$ in these operations.  As
a figure of merit of our entangling operation, we can prepare 3 (5) qubits in a
GHZ state with 98\% (95\%) fidelity~\cite{monz-superdecoherence}.  These
collective operations form a universal set of gates when used in conjuction
with single-qubit rotations $U_{Z_i}(\theta)=\exp(-i\frac{\theta}{2}Z_i)$,
which are realized by an off-resonant laser beam that can be adjusted to focus
on any ion.

For engineering dissipation, the key element of the mapping steps, shown as (i)
and (iii) in Box 1, is a single MS operation.  The two-qubit gate, step (ii),
is realized by a combination of collective and single-qubit operations.  The
dissipative mechanism, step (iv), is here carried out on the ancilla qubit by a
reinitialization into $|1\rangle$, as shown in
Fig.~\ref{fig:FigureExperiment}b.  Another dissipative
process~\cite{schindler-qec} can be used to prepare the system qubits in a
completely mixed state by the transfer
$|0\rangle\rightarrow(|0\rangle+|S'\rangle)/\sqrt{2}$ followed by optical
pumping of $|S'\rangle$ into $|1\rangle$, where $|S'\rangle$ is the electronic
level $S_{1/2}(m=1/2)$.

Qubit read-out is accomplished by fluorescence detection on the
$S_{1/2}\leftrightarrow P_{1/2}$ transition.  The ancilla qubit can be measured
without affecting the system qubits by applying hiding pulses that shelve the
system qubits in the $D_{5/2}$ state manifold during fluoresence
detection~\cite{roos-science-304-1478}.

We use these tools to implement up to three Bell-state cooling cycles on a
string of 2+1 ions.  Starting with the two system qubits in a completely mixed
state, we cool towards the Bell state $\ket{\Psi^-}$.  Each cooling cycle is
accomplished with a sequence of 8 entangling operations, 4 collective
unitaries and 6 single-qubit operations; see the Supplementary Information.
The cooling dynamics are probed by quantum state tomography of the system
qubits after every half cycle.  The reconstructed states are then used to map
the evolution of the Bell-state populations.

\begin{figure}
\iffigs\includegraphics{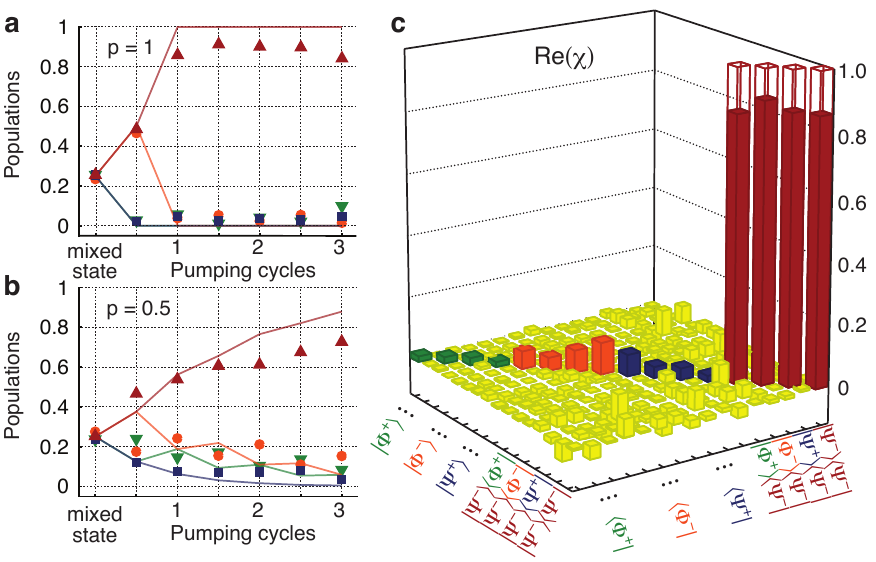}\fi
\caption{\textbf{Experimental signatures of Bell-state cooling.}  Evolution of
  the Bell-state populations $|\Phi^+\rangle$ (down triangles),
  $|\Phi^-\rangle$ (circles), $|\Psi^+\rangle$ (squares) and $|\Psi^-\rangle$
  (up triangles) of an initially mixed state under a cooling process with
  probability \textbf{a,} $p=1$ or deterministic and \textbf{b,} $p=0.5$.
  Error bars, not shown, are smaller than 2\% ($1\sigma$). \textbf{c,}
  Reconstructed process matrix $\chi$ (real part), displayed in the Bell-state
  basis, describing the deterministic cooling of the two ions after one and a
  half cycles.  The ideal process mapping any input state into the state
  $|\Psi^-\rangle$ has as non-zero elements only the four transparent bars
  shown.  The imaginary elements of $\chi$, ideally all zero, have an average
  magnitude of 0.004 and a maximum of 0.03.  The uncertainties in the elements
  of process matrix are smaller than 0.01 ($1\sigma$).
  \label{fig:bsc}}
\end{figure}

In a first experiment, we set the pumping probability at $p=1$ to observe
deterministic cooling, and we obtain the Bell-state populations shown in
Fig.~\ref{fig:bsc}a.  As expected, the system reaches the target state after
the first cooling cycle.  Regardless of experimental imperfections, the target
state population is preserved under the repeated application of further cooling
cycles and reaches up to 91(1)\% after 1.5 cycles (ideally 100\%).  In a second
experiment towards the simulation of master-equation dynamics, the probability
is set at $p=0.5$ to probe probabilistic cooling dynamics.  The target state is
then approached asymptotically (Fig.~\ref{fig:bsc}b).  After cooling the system
for 3 cycles with $p=0.5$, up to 73(1)\% of the initially mixed population
cools into the target state (ideally 88\%).  In order to completely
characterize the Bell-state cooling process, we also perform a quantum process
tomography~\cite{nielsen-book}.  As an example, the reconstructed process
matrix for $p=1$ after 1.5 cycles (Fig.~\ref{fig:bsc}c) has a Jamiolkowski
process fidelity~\cite{gilchrist-pra-71-062310} of 87.0(7)\% with the ideal
dissipative process $\rho_\text{S}\mapsto\ket{\Psi^-}\bra{\Psi^-}$ which maps
an arbitrary state of the system into the Bell state $\ket{\Psi^-}$.

\section{Four-Qubit Stabilizer Pumping}
\label{4stabilizer}

\begin{figure*}[t]
\iffigs\includegraphics{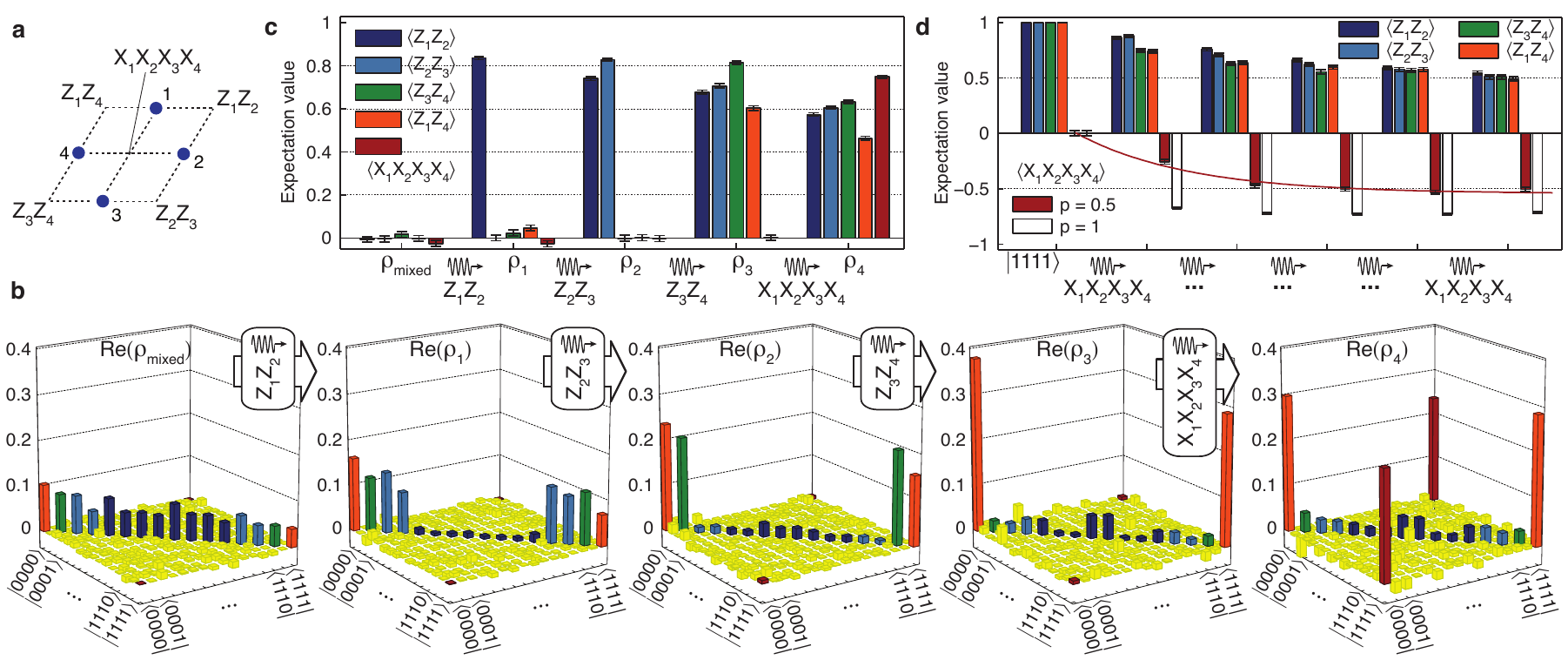}\fi
\caption{\label{fig:FigurePlaquette}\textbf{Experimental signatures of four-qubit stabilizer pumping.}  
\textbf{a,} Schematic of the four system qubits to be
  cooled into the GHZ state $(\ket{0000} + \ket{1111})/\sqrt{2}$, which is uniquely characterized as the simultaneous eigenstate with eigenvalue +1 of the shown stabilizers.  \textbf{b,} Reconstructed density matrices (real part) of the
  initial mixed state $\rho_\text{mixed}$ and subsequent states
  $\rho_{1,2,3,4}$ after sequentially pumping the stabilizers $Z_1Z_2$,
  $Z_2Z_3$, $Z_3Z_4$ and $X_1X_2X_3X_4$.  Populations in the initial mixed
  state with qubits $i$ and $j$ antiparallel, or in the -1 eigenspace of the
  $Z_iZ_j$ stabilizer, disappear after pumping this stabilizer into the +1
  eigenspace.  For example, populations in dark blue dissappear after
  $Z_1Z_2$-stabilizer pumping.  A final pumping of the stabilizer
  $X_1X_2X_3X_4$ builds up the coherence between $\ket{0000}$ and $\ket{1111}$,
  shown as red bars in the density matrix of $\rho_4$.  \textbf{c,} Measured
  expectation values of the relevant stabilizers; ideally, non-zero expectation
  values have a value of +1.  \textbf{d,} Evolution of the measured expectation
  values of the relevant stabilizers for repetitively pumping an initial state
  $|1111\rangle$ with probability $p=0.5$ into the -1 eigenspace of the
  stabilizer $X_1X_2X_3X_4$.  The incremental cooling is evident by the red
  line fitted to the pumped stabilizer expectation value.  The evolution of the
  expectation value $\langle X_1X_2X_3X_4\rangle$ for deterministic cooling
  ($p=1$) is also shown.  The observed decay of $\langle Z_iZ_j\rangle$ is due
  to imperfections and detrimental to the pumping process (see Supplementary
  Information).  Error bars in \textbf{c} and \textbf{d}, $\pm1\sigma$.}
\end{figure*}

The engineering of the system-environment coupling, as demonstrated by
Bell-state cooling above, can be readily extended to larger $n$-qubit open
quantum systems.  We illustrate such an engineering experimentally with the
dissipative preparation of a four-qubit Greenberger-Horne-Zeilinger (GHZ) state
$(\ket{0000} + \ket{1111})/\sqrt{2}$.  This state is uniquely characterized as
the simultaneous eigenstate of the four stabilizers $Z_1Z_2$, $Z_2Z_3$,
$Z_3Z_4$ and $X_1X_2X_3X_4$, all with eigenvalue +1 (see
Fig.~\ref{fig:FigurePlaquette}a).  Therefore, cooling dynamics into the GHZ
state are realized by four consecutive dissipative steps, each pumping the
system into the +1 eigenspaces of the four stabilizers.  In a system of 4+1
ions, we implement such cooling dynamics in analogy with the Bell-state cooling
sequence.  Here, however, the circuit decomposition of one cooling cycle
involves 16 five-ion entangling operations, 20 collective unitaries and 34
single-qubit operations; further details in the Supplementary Information.

In order to observe this deterministic cooling process into the GHZ state, we
begin by preparing the system ions in a completely mixed state.  The evolution
of the state of the system after each pumping step is characterized by quantum
state tomography.  The reconstructed density matrices shown in
Fig.~\ref{fig:FigurePlaquette}b for the initial and subsequent states arising
in each step have a fidelity, or state overlap~\cite{jozsa-jmo-41-2315}, with
the expected states of \{79(2),89(1),79.7(7),70.0(7),55.8(4)\}\%; see 
Supplementary Information for further details.  Since the
final state has a fidelity with the target GHZ state greater than 50\%, the
initially mixed state is cooled into a genuinely four-particle entangled
state~\cite{sackett-nat-404-256}.  The pumping dynamics is clearly reflected by
the measured expectation values of the stabilizers $Z_iZ_j$ ($ij=12,23,34,14$)
and $X_1X_2X_3X_4$ at each step, as shown in Fig.~\ref{fig:FigurePlaquette}c.

Although the simulation of a master equation requires small pumping
probabilities, as an exploratory study, we implement up to five consecutive
$X_1X_2X_3X_4$-stabilizer pumping steps with two probabilities $p=1$ and 0.5,
for the initial state $|1111\rangle$.  The measured expectation values of all
relevant stabilizers for pumping with $p=1$ are shown in
Fig.~\ref{fig:FigurePlaquette}d.  After the first step, the stabilizer
$X_1X_2X_3X_4$ reaches an expectation value of -0.68(1); after the second step
and up to the fifth step, it is preserved at -0.72(1) regardless of
experimental imperfections.

For $X_1X_2X_3X_4$-stabilizer pumping with $p=0.5$, the four-qubit expectation
value increases at each step and asymptotically approaches -0.54(1) (ideally
-1, fit shown in Fig.~\ref{fig:FigurePlaquette}d).  A state tomography after
each pumping step yields fidelities with the expected GHZ-state of \{53(1),
50(1), 49(1), 44(1), 41(1)\}\%.  From the reconstructed density matrices we determine that
the states generated after one to three cycles are genuinely multi-partite
entangled~\cite{guhne-njp-12-053002}.

\section{Coherent Four-Particle Interactions}
\label{sec:cs}

The coupling of the system to an ancilla particle, as used above for the
engineering of dissipative dynamics, can also be harnessed to mediate effective
coherent $n$-body interactions between the system qubits
\cite{nielsen-book,duer-pra-78-052325}.  The demonstration of a toolbox for
open-system quantum simulation is thus complemented by adding unitary maps
$\rho_{S}\mapsto U_S\rho_{S}U_S^{\dagger}$ to the dissipative elements
described above. Here, $U_S = \exp(-i \tau H_S)$ is the unitary time evolution
operator for a time step $\tau$, which is generated by a system Hamiltonian
$H_S$. In contrast to the recent achievements
\cite{friedenauer-nphys-4-757,kim-nature-465-590} of small-scale analog quantum
simulators based on trapped ions, where two-body spin Hamiltonians have been
engineered directly \cite{porras-prl-92-207901}, here we pursue a gate-based
implementation following the concept of Lloyd's digital quantum
simulator~\cite{lloyd-science-273-1073}, where the time evolution is decomposed
into a sequence of coherent (and dissipative) steps.

In particular, the available gate operations enable an \emph{experimentally
  efficient} simulation of $n$-body spin interactions \cite{mueller-2010},
which we illustrate by implementing time dynamics of a four-body Hamiltonian
$H_S =g X_1X_2X_3X_4$. This example is motivated by the efforts to
experimentally realize Kitaev's toric code Hamiltonian
\cite{Kitaev-annalsphys-303-2}, which is a sum of commuting four-qubit
stabilizer operators representing four-body spin interactions. This
paradigmatic model belongs to a whole class of spin systems, which have been
discussed in the context of topological quantum
computing~\cite{nayak-revmodphys-80-1083} and quantum phases exhibiting
topological order~\cite{moessner-prl-86-1881}.

\begin{figure}[ht]
\iffigs\includegraphics{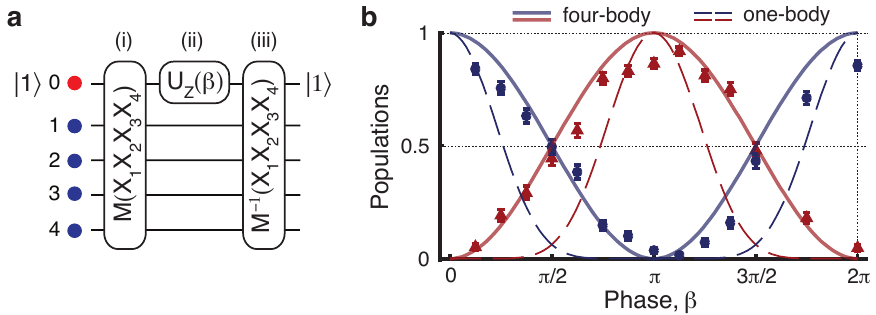}\fi
\caption{\label{fig:cs} \textbf{Coherent simulation of $4$-body spin
    interactions.}  \textbf{a,} The elementary building block for the
  simulation of coherent evolution $U_S = \exp(-i \tau H_S)$ corresponding to
  the four-body Hamiltonian $H_S = g X_1X_2X_3X_4$ is implemented by a circuit
  of three operations: (i) First, a $5$-qubit operation
  $M(X_1X_2X_3X_4)$, here realized by a single entangling 5-ion MS gate
  $U_{S_x^2}(\pi/2)$, coherently maps the information, whether the four system
  spins are in the +1(-1) eigenspace of $X_1X_2X_3X_4$ onto the internal states
  $\ket{0}$ and $\ket{1}$ of the ancilla qubit.  (ii) Due to this mapping, all
  +1 (-1) eigenstates of $X_1X_2X_3X_4$ acquire a phase $\beta/2$ ($-\beta/2$)
  by the single-qubit rotation $U_{Z}(\beta)$ on the ancilla ion. (iii) After
  the mapping is inverted, the ancilla qubit returns to its initial state
  $\ket{1}$ and decouples from the four system qubits, which in turn have
  evolved according to $U_S$. The simulation time step $\tau$ is related to the
  phase by $\beta = 2 g \tau$. \textbf{b,} Experimentally measured populations
  in state $|0000\rangle$ (up triangles) and $|1111\rangle$ (circles) as a
  function of $\beta$ for a single application of $U_S$ to the initial state
  $\ket{1111}$ of the four system qubits (error bars, $\pm 1 \sigma$).  The
  solid lines show the ideal behavior.  For comparison, the dashes lines
  indicate these populations for simultaneous single-qubit (one-body)
  oscillations, each driven by the rotation $\exp(-i \frac{\beta}{2} X_i)$. }
\end{figure}

The elementary unitary operation $U_S$ can be decomposed into a compact
sequence of three coherent operations, as explained in Fig.~\ref{fig:cs}a. In an
experiment carried out with 4+1 ions, we apply $U_S$ for different
values of $\tau$ to the system ions initially prepared in $\ket{1111}$. We
observed coherent oscillations in the subspace spanned by $\ket{0000}$ and
$\ket{1111}$, as shown in Fig.~\ref{fig:cs}b.  We characterize our
implementation of $U_S$ by comparing the expected and measured states,
determined by quantum state tomography, for each value of $\tau$.  The fidelity
between the expected and measured states is on average 85(2)\%.

\section{QND Measurement of Four-Qubit Stabilizer Operators}
\label{sec:QND}

Our toolbox for quantum simulation of open systems is extended by the
possibility of reading out $n$-body observables in a nondestructive way, which
we illustrate here for a $4$-qubit stabilizer operator $X_1X_2X_3X_4$.  As
above, we first coherently map the information about whether the system spins
are in the +1(-1) eigenspace of the stabilizer operator onto the logical states
$\ket{0}$ and $\ket{1}$ of the ancilla qubit. In contrast to the engineering of
coherent and dissipative maps above, where this step was followed by single-and
two-qubit gate operations, here we proceed instead by measuring the ancilla
qubit.  

Thus, depending on the measurement outcome for the ancilla, the system qubits
are projected onto the corresponding eigenspace of the stabilizer: $\rho_S
\mapsto P_+ \rho_S P_+/N_+$ $(P_-\rho_S P_-/N_-)$ for finding the ancilla in
$\ket{0}$ ($\ket{1}$) with the normalization factor $N_\pm=\text{Tr}(P_\pm\rho_S
P_\pm)$.  Here, $P_\pm = \frac 12 (1 \pm X_1X_2X_3X_4)$ denote the projectors
onto the $\pm 1$ eigenspaces of the stabilizer operator. Note that our
measurement is QND in the sense that (superposition) states within one of the
two eigenspaces are not affected by the measurement.

In the experiment with 4+1 ions, we prepare different four-qubit system input
states (tomographically characterize in additional experiments), carry out the
QND measurement and tomographically determine the resulting system output
states.

To characterize how well the measurement device prepares a definite state, we
use as input $\ket{1111}$, which is a non-eigenstate of the stabilizer.  In
this case, when the ancilla qubit is found in $|0\rangle$ or $|1\rangle$ the
system qubits are prepared in the state $(\ket{0000} \pm \ket{1111})/\sqrt{2}$
by the QND measurement.  Experimentally we observe this behaviour with a
quantum state preparation (QSP) fidelity~\cite{ralph-pra-73-012113} of
$F_\text{QSP} = 73(1)\%$.  On the other hand, for a stabilizer eigenstate, the
QND measurement preserves the stabilizer expectation value.  Experimentally,
for the input state $(|0011\rangle-|1100\rangle)/\sqrt{2}$, we observe a QND
fidelity~\cite{ralph-pra-73-012113} of $F_\text{QND}=96.9(6)\%$.  For more
details see the Supplementary Information.

Our measurement of $n$-body observables is an essential ingredient in quantum
error correction and quantum computing protocols. In contrast to the
\textit{open-loop} experiments presented here \cite{lloyd-pra-65-010101}, this ability also enables an
alternative approach for system-environment engineering: The outcome from
measurements of the environment can be classically processed and used for
feedback operations on the system. This procedure paves the way to
\textit{closed-loop} simulation scenarios in open quantum systems.

\section{Outlook}

Our experimental demonstration of a toolbox of elementary building blocks in a
system of trapped ions should be seen as a first, and conceptual step towards
the realization of an open quantum system simulator, with dynamics governed by
the interplay of coherent and dissipative evolution.  Such a quantum device has
applications in various fields~\cite{buluta-science-326-108} including
condensed-matter physics and quantum chemistry, and possibly in modelling
quantum effects in biology~\cite{qbio}. In addition to quantum simulation, it
enables alternative approaches to quantum computing
\cite{verstraete-nphys-5-633}.

Although the present experiments were performed with a linear ion-trap quantum
computer architecture, the ongoing development of two-dimensional trap
arrays~\cite{schmied-prl-102-233002} promises scalable implementations of
Kitaev's toric code~\cite{Kitaev-annalsphys-303-2} and related spin models, as
discussed in the context of topological quantum computing.  Following our
original proposal~\cite{weimer-nphys-6-382}, these ideas can be realized with
neutral atoms in optical lattices and can be easily adapted to other physical
platforms ranging from optical, atomic and molecular systems to solid-state
devices.

\section{Acknowledgments}
We would like to thank K. Hammerer, I. Chuang, and O. G\"uhne for discussions
and T. Northup for critically reading the manuscript.  We gratefully
acknowledge support by the Austrian Science Fund (FOQUS), the European
Commission (AQUTE), the Institut f\"ur Quanteninformation GmbH, and a Marie
Curie International Incoming Fellowship within the 7th European Community
Framework Programme.

\section{Author Contributions}

M.M. and J.T.B. developed the research, based on theoretical ideas proposed
originally by P.Z.; J.T.B., P.S. and D.N. carried out the experiment; J.T.B.,
P.S. and T.M. analysed the data; P.S., J.T.B., D.N., T.M., M.C., M.H. and
R.B. contributed to the experimental setup; M.M., J.T.B.  and P.Z. wrote the
manuscript, with revisions provided by C.F.R.; all authors contributed to the
discussion of the results and manuscript.

\bibliography{cooling}

\bibliographystyle{mynaturelikestyle}

\end{document}


\title{An Open-System Quantum Simulator with Trapped Ions\\
SUPPLEMENTARY INFORMATION}

\author{Julio T. Barreiro*$^{,1}$, Markus M\"uller*$^{,2,3}$,
 Philipp Schindler$^1$, Daniel Nigg$^1$, Thomas Monz$^1$, 
 Michael Chwalla$^{1,2}$, Markus Hennrich$^1$, Christian Roos$^{1,2}$, 
 Peter Zoller$^{2,3}$ and Rainer Blatt$^{1,2}$\\[2mm]
$^{1}$Institut f\"ur Experimentalphysik, Universit\"at Innsbruck,
  Technikerstrasse 25, 6020 Innsbruck, Austria\\
$^{2}$Institut f\"ur Quantenoptik und Quanteninformation,
\"Osterreichische Akademie der Wissenschaften,\\ \hspace{2mm}Technikerstrasse 21A, 6020 Innsbruck,
  Austria\\
$^{3}$Institut f\"ur Theoretische Physik,
  Universit\"at Innsbruck, Technikerstrasse 25, 6020 Innsbruck,
  Austria\\
* These authors contributed equally to this work.}

\maketitle

\tableofcontents
\vspace{5mm}
\noindent\textsf{\textbf{\small List of tables}}\\

\noindent{{\footnotesize
\begin{tabular}[b]{rp{75.3mm}@{\hspace{3mm}}r}
\textbf{I}:& QND probability distributions with hiding &{\raggedright5}\\[5mm]
\textbf{II}:& QND probability distributions without hiding &5\\[5mm] 
\textbf{III}:& QND fidelities with hiding &6\\[5mm]
\textbf{IV}:& QND fidelities without hiding &7\\[5mm] 
\end{tabular}
}}

\section{Bell-state cooling}

\subsection{Implemented Kraus maps}

The Bell state $\ket{\Psi^-}$ is not only uniquely determined as the
simultaneous eigenstate with eigenvalue -1 of the two stabilizer operators
$X_1X_2$ and $Z_1Z_2$ (as mentioned in the text), but also by $X_1X_2$ and
$Y_1Y_2$. In the experiment, we implemented cooling into $\ket{\Psi^-}$ by
engineering the two Kraus maps $\rho_S \mapsto E_1\rho_S E_1^\dagger +
E_2\rho_S E_2^\dagger$ and $\rho_S \mapsto E'_1\rho_S {E'_1}^{\dagger} +
E'_2\rho_S {E'_2}^{\dagger}$, where
\begin{align}
E_{1} & = \sqrt{p} \, Y_1 \frac{1}{2} \left(1 + X_1X_2 \right), \nonumber \\
E_{2} & = \frac{1}{2} \left( 1 - X_1X_2 \right) +  \sqrt{1-p} \,\frac{1}{2} \left( 1 + X_1X_2 \right),\nonumber\\
E'_{1} & = \sqrt{p} \, X_1 \frac{1}{2} \left(1 +Y_1Y_2 \right), \nonumber \\
E'_{2} & =  \frac{1}{2} \left( 1 - Y_1Y_2 \right) +  \sqrt{1-p} \,\frac{1}{2} \left( 1 + Y_1Y_2 \right),\nonumber
\end{align}
which generate pumping into the -1 eigenspaces of $X_1X_2$ and $Y_1Y_2$
(instead of pumping into the eigenspaces of $X_1X_2$ and $Z_1Z_2$ as explained
in Box~1 of the main text). The reason for pumping into the eigenspaces of
$X_1X_2$ and $Y_1Y_2$ is that the mapping and unmapping steps, shown as (i) and
(iii) in Box 1, are realized by a single MS gate $U_{X^2}(\pi/2)$ and
$U_{Y^2}(\pi/2)$, respectively.

\subsection{Circuit decomposition}

The map for pumping into the -1 eigenspace of $X_1X_2$ was implemented by the
unitary 
\begin{equation}
U_{X^2(\pi/2)} \,C(p) \, U_{X^2(\pi/2)}
\end{equation}
(corresponding to steps (i) - (iii) in Box~1) followed by an optical pumping of
the ancilla qubit to $\ket{1}$.  Here, the two-qubit controlled gate is
\begin{eqnarray}
C(p) & = & \ket{0}\bra{0}_0 \otimes \exp(i\alpha Z_1) + \ket{1}\bra{1}_0
\otimes \openone \nonumber \\ & = & \exp \left[ \frac{1}{2} (1+ Z_0) i \alpha Z_1
  \right] \nonumber \\ & = & U_{Z_1} (-\alpha)\, U_{Y}(\pi/2)\,
U_{X^2}^{(0,1)}(-\alpha) \,U_{Y}(-\pi/2)
\end{eqnarray}
where $U_{X^2}^{(0,1)}(-\alpha) = \exp(i(\alpha/2) X_0 X_1)$ denotes an MS gate
acting only on the ancilla and the first system qubit. This two-qubit MS gate
operation was implemented in the experiment by the use of refocusing
techniques~\citep{nebendahl-pra-79-012312}.  In more detail, the gate
$U_{X^2}^{(0,1)}$ was realized by interspersing two of the available
three-qubit MS gate operations with single-ion light shifts on the second
system qubit which induces a $\pi$-phase shift between the qubit states.
Alternatively, this refocusing could be avoided, and the sequences further
simplified, by hiding the population of individual ions (here the second system
ion) which are not supposed to participate in collective coherent operations in
electronic levels decoupled from the driving laser excitation. More details on
how to systematically decompose Kraus maps into the experimentally available
ion-trap gate operations, in particular the multi-ion MS entangling gate, can
be found in~\cite{mueller-2010}.

\begin{figure}[h!]
\begin{center}
\iffigs \includegraphics{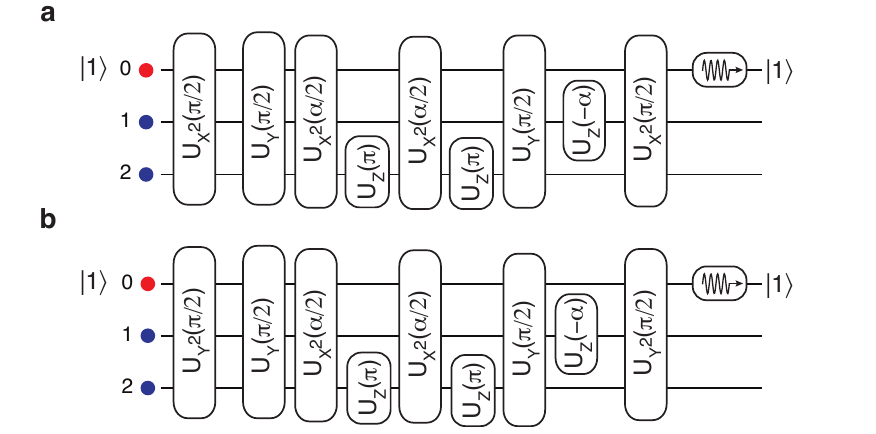}\fi
\end{center}
\caption{\label{fig:bsc-circuits} \textbf{Experimental sequences for Bell-state
    cooling.} Pumping into the eigenspaces of eigenvalue -1 of $X_1X_2$ (circuit \textbf{a,}) and
  $Y_1Y_2$ (circuit \textbf{b,}) occurs with a probability $p$ in each step, where $\sin^2\alpha =
  p$.}
\end{figure}

The circuit decompositions for the experimental implementation of the two maps
are shown in Fig.~\ref{fig:bsc-circuits}.  We note that the circuits have
been simplified at the expense of implementing in addition in each dissipative
map a flip operation $Y_1Y_2$ on the two system qubits. However, as this
additional unitary corresponds to one of the stabilizers into whose -1
eigenspace cooling is performed, this does not interfere with the cooling
dynamics.

Cooling with unit pumping probability $p=1$ corresponds to $\alpha = \pi/2$,
whereas $p=0.5$ is realized with by setting $\alpha = \pi/4$. In the
experiment, the "fundamental" MS gate was calibrated to implement
$U_{X^2}(\alpha/2)$. The fully entangling operation $U_{X^2}(\pi/2)$ at the
beginning and the end of the sequence Fig.~\ref{fig:bsc-circuits}a was then
implemented by applying the $U_{X^2}(\alpha/2)$ operation twice (for $p=1$) or
four times (for $p=0.5$).  The fully entangling operations $U_{Y^2}(\pi/2)$ in
Fig.~\ref{fig:bsc-circuits}b were implemented by two- and four-fold application
of the "fundamental" MS gate with a shifted optical phase of the driving laser
(cf.~Section 2 in the main text).

\subsection{Additional data}

The initially mixed state was prepared with a fidelity of F=99.6(3)\% with
respect to the ideal state $\frac{1}{4}\openone_{4\times4}$.

Physical process matrices were reconstructed with maximum likelihood
techniques~\citep{jezek-pra-68-012305}.  An error analysis was carried out via
Monte Carlo simulations over the multinomially distributed measurement outcomes
of the state and process tomography.  For each process and state, 200 Monte
Carlo samples were generated and reconstructed via maximum-likelihood
estimation.

\section{Four-qubit stabilizer pumping}

Expectation values of the stabilizer operators $Z_1Z_2$, $Z_2Z_3$, $Z_3Z_4$ and
$X_1X_2X_3X_4$ were not determined from the reconstructed density matrices of
the system qubits.  Instead, we performed fluorescence measurements in the $X$
and $Z$ basis on 5250 copies of the corresponding quantum states (for $p=0.5$
cooling, 2100 copies were measured).  The error bars were then determined from
the multinomially distributed raw data.

\subsection{Cooling}
\label{sec:Cooling}

Cooling into the GHZ state $(\ket{0000}+\ket{1111})/\sqrt{2}$ was realized by a
pumping cycle where the four system qubits were deterministically pumped into
the +1 eigenspaces of the stabilizers $Z_1Z_2$, $Z_2Z_3$, $Z_3Z_4$ and
$X_1X_2X_3X_4$.

Pumping into the -1 eigenspace of $Z_1Z_2$ in the first cooling step could be
achieved in complete analogy with Bell state cooling, i.e.~by implementing a
dissipative map, which only involves operations on the ancilla qubit and the
system qubits \#1 and \#2, whereas the system qubits \#3 and \#4 remain
completely unaffected.  This could either be achieved through refocusing
techniques or by hiding system ions \#3 and \#4 in electronically decoupled
states for the duration of the dissipative circuit.

\begin{figure}[h!]
\begin{center}
\iffigs \includegraphics{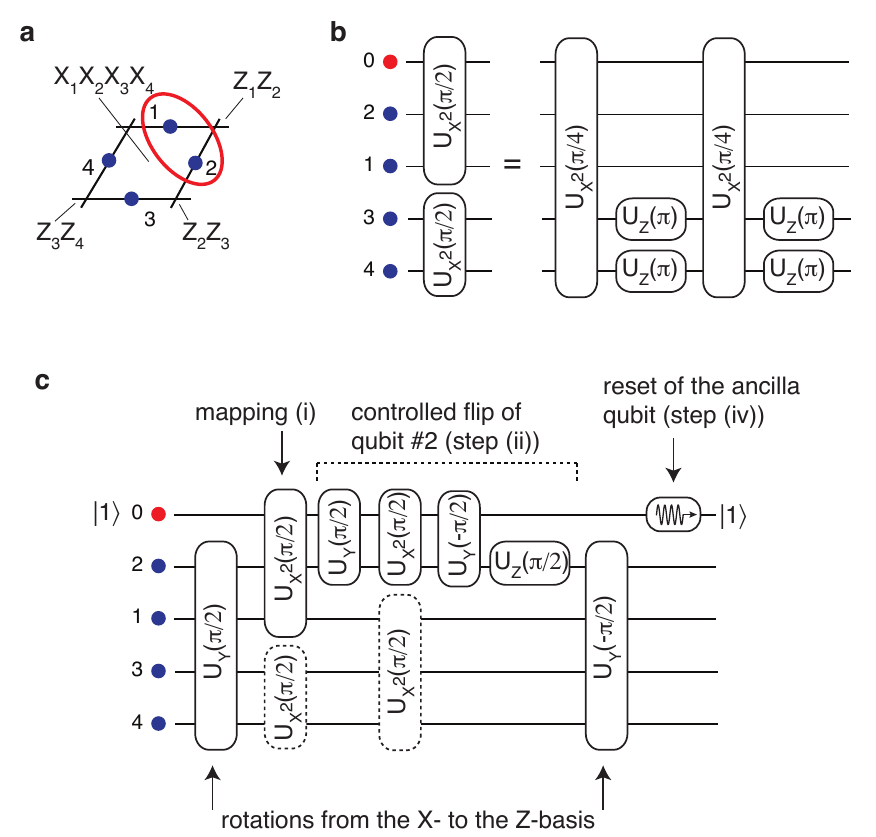}\fi
\end{center}
\caption{\label{fig:Z1Z2cooling} \textbf{Pumping into the +1 eigenspace of the
    $Z_1Z_2$ stabilizer operator} \textbf{a,} Ideally, only the ancilla qubit
  and the two system qubits \#1 and \#2 are involved in the
  circuit. \textbf{b,} An entangling gate acting on these three ions can be achieved by 
  a refocusing technique, where ions \#3 and \#4 decouple from the dynamics. However, 
  the latter ions still become entangled. However, these residual interactions are not harmful to the cooling,
  as they do not affect the expectation values of the other two-qubit stabilizer operators. 
  \textbf{c,} Dashed operations in the quantum circuit indicate such residual entangling operations.}
\end{figure}

In the experiment, however, we used a few simplications, which are
schematically shown in Fig.~\ref{fig:Z1Z2cooling} and listed below:

\begin{itemize}
\item For deterministic cooling ($p=1$), the inverse mapping step (shown in Box 1)
is not necessary and has been taken out.

\item In the coherent mapping step (shown in Box 1) the information about
  whether the system ions are in a $\pm 1$ eigenstate of $Z_1Z_2$ is mapped
  onto the logical states of the ancilla qubit. This step ideally only involves
  the ancilla and the system qubits \#1 and \#2. One way to achieve this
  three-qubit operation without affecting the system qubits \#3 and \#4, is to
  combine the available five-ion MS gate with appropriately chosen refocusing
  pulses, i.e.~light shift operations on individual ions. Those would have to
  be chosen such that ions \#0, \#1 and \#2 become decoupled from ions \#3 and
  \#4, and furthermore residual interactions between ions \#3 and \#4 cancel
  out. However, it turns out that residual interactions between ions \#3 and
  \#4 can be tolerated: although not required for the $Z_1Z_2$-pumping
  dynamics, they are not harmful, as they do not alter the expectation values
  of the other two-qubit stabilizers $Z_2Z_3$ and $Z_3Z_4$. In our experiment
  the decoupling of ions \#0, \#1 and \#2 from the ions \#3 and \#4 was
  achieved by the circuit shown in Fig.~\ref{fig:Z1Z2cooling}b.

The additional interactions in the pumping of the two-qubit stabilizer
operators $Z_iZ_j$ affect the state of the system qubits with respect to the
four-qubit stabilizer $X_1X_2X_3X_4$. However, this effect is not detrimental
to the cooling, provided the pumping into the eigenspace of $X_1X_2X_3X_4$ is
performed as the final step in the cooling cycle.

\item In the employed sequence, the number of single-qubit rotations was
  reduced wherever possible. Essential single-qubit light shift operations,
  such as those needed for re-focusing operations, were kept.

\item Local rotations of the system ions at the end of a cooling step, which
  would be compensated at the beginning of the subsequent cooling step, were
  omitted when several dissipative maps were applied in a row. The
  corresponding gates of the sequences are displayed in blue in Steps 1-3.
\end{itemize}

These simplifications allowed us to significantly reduce the length and
complexity of the employed gate sequences for one stabilizer pumping step.  The
compressed gate sequences as used in the experiment are explicitly given below.

Step 1 (pumping into the +1 eigenspace of $Z_1Z_2$):
\begin{eqnarray*}
{\color{blue}U_Y(-\pi/2)}\,
U_{Z_2}(-\pi/2)\\
U_X    ( \pi/2)\,
U_{Z_2}(-\pi/2)\,
U_X    (-\pi/2)\\
U_{Z_1}( \pi)\,
U_{X^2}( \pi/4)\,
U_{Z_2}( \pi)\,
U_{Z_0}( \pi)\,
U_{X^2}( \pi/4)\\
U_X    (-\pi/2)\,
U_{Z_2}(-\pi/2)\,
U_{Z_0}(-\pi/2)\,
U_X    ( \pi/2)\\
U_{X^2}( \pi/4)\,
U_{Z_4}( \pi)\,
U_{Z_3}( \pi)\,
U_{X^2}( \pi/4)\\
U_Y    ( \pi/2)\,
U_X    (-\pi/2)\,
U_{Z_0}(-\pi/2)\,
U_X    ( \pi/2)\,
\end{eqnarray*}

Step 2 (pumping into the +1 eigenspace of $Z_2Z_3$):
\begin{eqnarray*}
{\color{blue}U_Y(-\pi/2)}\,
{\color{red}U_{Z_3}(-\pi/2)}\\
U_X    ( \pi/2)\,
U_{Z_3}(-\pi/2)\,
U_X    (-\pi/2)\\
U_{Z_2}( \pi)\,
U_{X^2}( \pi/4)\,
U_{Z_3}( \pi)\,
U_{Z_0}( \pi)\,
U_{X^2}( \pi/4)\\
U_X    (-\pi/2)\,
U_{Z_3}(-\pi/2)\,
U_{Z_0}(-\pi/2)\,
U_X    ( \pi/2)\\
U_{X^2}( \pi/4)\,
U_{Z_4}( \pi)\,
U_{Z_1}( \pi)\,
U_{X^2}( \pi/4)\\
{\color{blue}U_Y    ( \pi/2)\,
U_X    (-\pi/2)\,
U_{Z_0}(-\pi/2)\,
U_X    ( \pi/2)\,}
\end{eqnarray*}

Step 3 (pumping into the +1 eigenspace of $Z_3Z_4$):
\begin{eqnarray*}
U_Y(-\pi/2)\,
{\color{red}U_{Z_4}(-\pi/2)}\\
U_X    ( \pi/2)\,
U_{Z_4}(-\pi/2)\,
U_X    (-\pi/2)\\
U_{Z_3}( \pi)\,
U_{X^2}( \pi/4)\,
U_{Z_4}( \pi)\,
U_{Z_0}( \pi)\,
U_{X^2}( \pi/4)\\
U_X    (-\pi/2)\,
U_{Z_4}(-\pi/2)\,
U_{Z_0}(-\pi/2)\,
U_X    ( \pi/2)\\
U_{X^2}( \pi/4)\,
U_{Z_2}( \pi)\,
U_{Z_1}( \pi)\,
U_{X^2}( \pi/4)\\
{\color{blue}U_Y    ( \pi/2)\,
U_X    (-\pi/2)\,
U_{Z_0}(-\pi/2)\,
U_X    ( \pi/2)\,}
\end{eqnarray*}

Step 4 (pumping into the +1 eigenspace of $X_1X_2X_3X_4$):
\begin{eqnarray*}
U_X    (-\pi/2)\\
{\color{red}U_{Z_4}(-\pi/2)\,}
U_X    ( \pi/2)\,
{\color{red}U_{Z_4}(-\pi/2)}\\
U_{X^2}( \pi/4)\,
U_{Z_4}( \pi)\,
U_{Z_0}( \pi)\,
U_{X^2}( \pi/4)\\
U_{Z_4}(-\pi/2)\,
U_X    (-\pi/2)\,
U_{Z_0}(-\pi/2)\,
U_X    ( \pi/2)\\
U_{X^2}( \pi/4)\,U_{X^2}( \pi/4)
\end{eqnarray*}

\begin{figure*}[h!]
\iffigs \begin{center}
\includegraphics{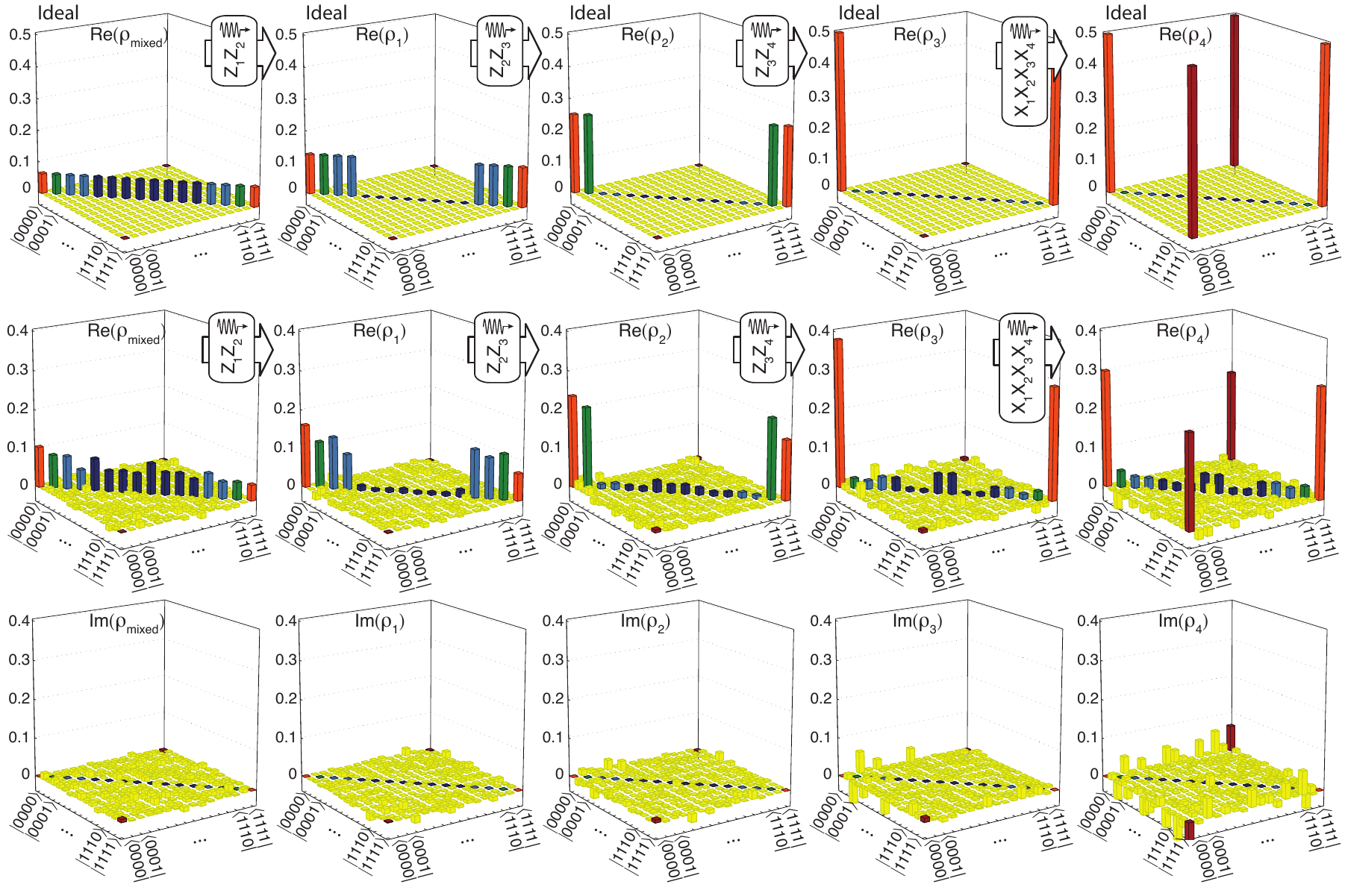}
\end{center}\fi
\caption{\label{fig:allrhos}\textbf{Ideal and reconstructed density matrices of plaquette
    cooling.}  An inital mixed state $\rho_\text{mixed}$ is sequentially pumped
  by the stabilizers $Z_1Z_2$, $Z_2Z_3$, $Z_3Z_4$ and $X_1X_2X_3X_4$ driving
  the system into the states $\rho_{1,2,3,4}$.}
\end{figure*}

Figure~\ref{fig:allrhos} shows the reconstructed density matrices (real and
imaginary parts) for every step of the cooling cycle.  The complete circuit
decomposition of one cooling cycle involves 16 five-ion entangling operations,
28 (20) collective unitaries and 36 (34) single-qubit operations with (without)
optional operations in blue.  The reset operation involves further pulses not
accounted for above.

\subsection{Repeated four-qubit stabilizer pumping}

\begin{figure}[h!]
\iffigs\begin{center}
\includegraphics{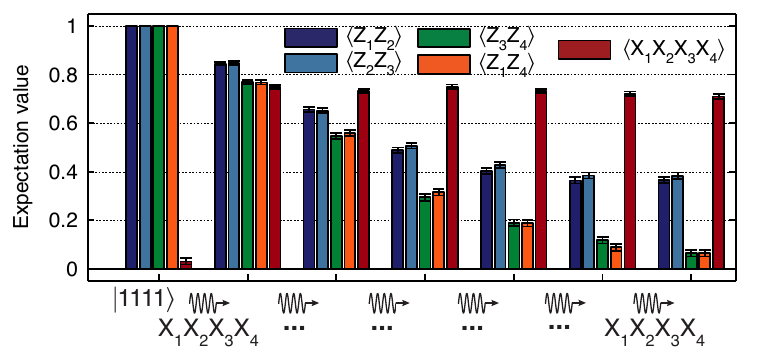}
\end{center}\fi
\caption{\label{fig:repeated}\textbf{Measured expectation value of stabilizers
    for repeated pumping without sequence optimization.}  The expectation
  values of $Z_1Z_4$ and $Z_3Z_4$ show a significantly faster decay than those
  for $Z_1Z_2$ and $Z_2Z_3$.  In every step of the cooling, most single-ion
  light-shift operations are applied to the fourth system qubit.}
\end{figure}

To study the robustness of the dissipative operation, we prepared the initial
state $\ket{1111}$ and subsequently applied repeatedly the dissipative map for
pumping into the +1 eigenspace of the four-qubit stabilizer $X_1X_2X_3X_4$. We
observed that after a single dissipative step a non-zero expectation value of
$X_1X_2X_3X_4$ built up and stayed constant under subsequent applications of
this dissipative map. However, due to imperfections in the gate operations, the
expectation values of the two-qubit stabilizers decreased, ideally they should
not be affected by the $X_1X_2X_3X_4$-pumping step (see
Fig.~\ref{fig:repeated}). Interestingly, the expectation values of $Z_1Z_4$ and
$Z_3Z_4$ decayed significantly faster than those for $Z_1Z_2$ and
$Z_2Z_3$. This decay can be explained by the fact that in the gate sequence
used for pumping into the +1 eigenspace of $X_1X_2X_3X_4$, step 4 above,
single-ion light-shift operations are applied only to the fourth system qubit
and the ancilla. This indicates that errors in the single-qubit gates applied
to the fourth system ion accumulate under the repeated application of the
dissipative step, and thus affect the stabilizers $Z_1Z_4$ and $Z_3Z_4$ which
involve this system qubit more strongly than the others. This destructive
effect can be minimized by alternating the roles of the system qubits.

\begin{figure}[h!]
\iffigs\begin{center}
\includegraphics{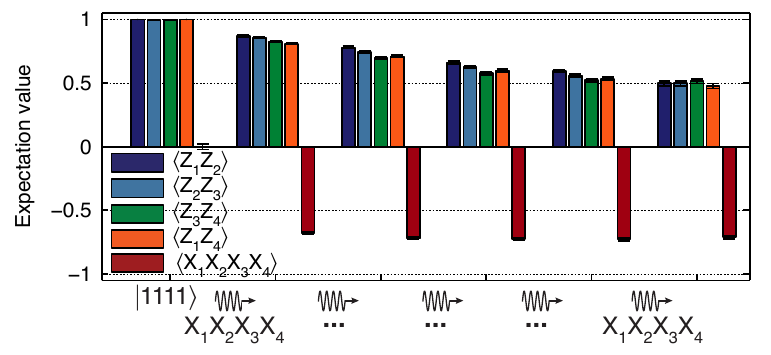}
\end{center}\fi
\caption{\label{fig:deterministic}\textbf{Measured expectation value of
    stabilizers for repeated pumping with sequence optimization.}  All
  two-qubit stabilizers decay at the same rate during cooling.  In step
  1,2,3,4, and 5 the single-qubit light-shift operations were applied on the
  system qubits 4,3,2,1, and 1, respectively.}
\end{figure}

Such optimization has been done for the dissipative dynamics shown in
Fig.~\ref{fig:deterministic}. Here, starting from the initial state
$\ket{1111}$, repeated pumping into the -1 eigenspace of $X_1X_2X_3X_4$ has
been implemented by the sequence

\begin{eqnarray*}
U_{X^2}( \pi/8)\,U_{X^2}( \pi/8)\,U_{X^2}( \pi/8)\,U_{X^2}( \pi/8)\\
U_X    (-\pi/2)\\
{\color{red}U_{Z_4}(-\pi/2\times {\color{blue}p})\,}
U_X    ( \pi/2)\,
{\color{red}U_{Z_4}( \pi  )}\\
U_{Y^2}( \pi/4\times {\color{blue}p})\,
U_{Z_0}( \pi  )\,
{\color{red}U_{Z_4}( \pi  )\,}
U_{Y^2}( \pi/4\times {\color{blue}p})\\
U_Y    ( \pi/2)\,
U_{Z_0}(-\pi/2)\,
U_Y    (-\pi/2)\\
U_{X^2}( \pi/8)\,U_{X^2}( \pi/8)\,U_{X^2}( \pi/8)\,U_{X^2}( \pi/8)
\end{eqnarray*}

Here, we observed that indeed the expectation values of all two-qubit
stabilizers decreased at the same pace and at a slightly slower rate (see
Fig.~\ref{fig:deterministic}).  Upon repeating the sequence above 1,2,3,4, and
5 times, we changed the operations shown in red to act on qubits 4,3,2,1, and
1, respectively.  The stabilizer expectation values for deterministic cooling,
or $p=1$, are shown in Fig.~\ref{fig:deterministic}.

\subsection{Pushing ``anyons'' around}
In Kitaev's toric code~\citep{Kitaev-annalsphys-303-2}, spins are located on
the edges of a two-dimensional square lattice. The Hamiltonian
\begin{equation}
H = - g (\sum_p A_p + \sum_v B_p) 
\end{equation}
is a sum of mutually commuting four-qubit stabilizers $A_p = \prod_{i \in p}
X_i$ and $B_v = \prod_{i \in v} Z_i$, which describe four-spin interactions
between spins located around plaquettes $p$ and vertices $v$ of the
lattice. The ground state of the Hamiltonian is the simultaneous +1 eigenstate
of all stabilizer operators. The model supports two types of excitations that
obey anyonic statistics under exchange (braiding), and they correspond to -1
eigenstates of either plaquette or vertex stabilizers.

For a minimal instance of this model, represented by a single plaquette of four
spins located on the edges, the Hamiltonian contains a single four-qubit
interaction term $X_1X_2X_3X_4$ and pairwise two-spin interactions $Z_iZ_j$ of
spins sharing a corner of the plaquette. The ground state as the simultaneous
+1 eigenstate of these stabilizers is the GHZ-state $(\ket{0000} +
\ket{1111})/\sqrt{2}$. States corresponding to -1 eigenvalues of a two-qubit
stabilizer $Z_iZ_j$ can be interpreted as a configuration with an excitation
located at the corner between the two spins $i$ and $j$. Similarly, a
four-qubit state with an eigenvalue of -1 with respect to $X_1X_2X_3X_4$, would
correspond to an anyonic excitation located at the center of the plaquette.

\begin{figure}[h!]
\iffigs\begin{center}
\includegraphics{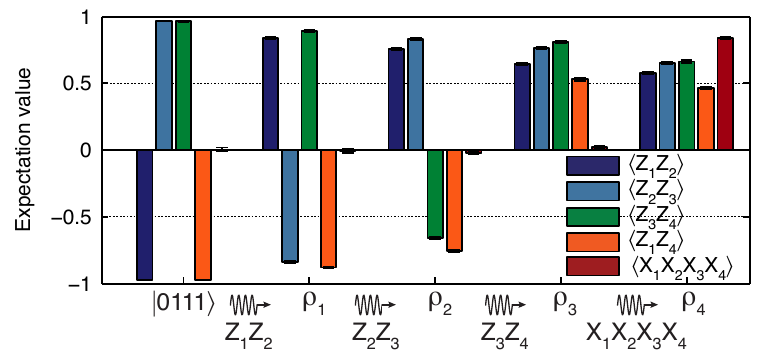}
\end{center}\fi
\caption{\label{fig:pushing}\textbf{Pushing ``anyons'' around by dissipation.}
  Measured expectation values of stabilizer operators for cooling dynamics of
  pumping into the +1 eigenspaces of $Z_1Z_2$, $Z_2Z_3$, $Z_3Z_4$ and
  $X_1X_2X_3X_4$, starting in the state $\ket{0111}$.}
\end{figure}

In the experiment we prepared an initial state $\ket{0111}$ and then performed
the cooling cycle of four deterministic pumping steps into the +1 eigenspaces
of $Z_1Z_2$, $Z_2Z_3$, $Z_3Z_4$ and $X_1X_2X_3X_4$, using the sequences for
Steps 1 to 4 given in section \ref{sec:Cooling}. The expectation values of the
stabilizer operators for the initial state and the four spins after each
pumping step are shown in Fig.~\ref{fig:pushing}. The dissipative dynamics can
be visualized as follows: For the initial state with $\langle Z_1Z_2\rangle =
-1 $ and $\langle Z_1Z_4\rangle = -1 $ a pair of excitations is located on the
upper left and right corners of the plaquette, whereas $\langle
X_1X_2X_3X_4\rangle = 0 $ implies an anyon of the other type is present at the
center of the plaquette with a probability 50\%. In the first cooling step,
where the first two spins are pumped into the +1 eigenspace of $Z_1Z_2$, the
anyon at the upper right corner is dissipatively pushed to the lower right
corner of the plaquette. In the third step of pumping into the +1 eigenspace of
$Z_3Z_4$, the two excitations located on the upper and lower lefts corners fuse
and disappear from the system. In the final step of pumping into the +1
eigenspace of $X_1X_2X_3X_4$, the anyon with a probability of 50\% at the
center of the plaquette is pushed out from the plaquette.

However, we'd like to stress that borrowing concepts from topological spin
models, such as anyonic excitations, here is merely a convenient language to
phrase and visualize the dissipative dynamics. In the present work with up to
five ions, we do \textit{not} explore the physics of topological spin models,
since (i) in a minimal system of four spins the concepts developed for larger
lattice models become questionable, and more importantly, (ii) during the
implemented cooling dynamics the underlying (four-body) Hamiltonian of the
model was not present. We rather demonstrate the basic tools which will allow
one to explore this physics once larger, two-dimensional systems become
available in the laboratory.   

We note that photon experiments have reported the observation of correlations
compatible with the manipulations of ``anyons'' in a setup representing two
plaquettes~\citep{lu-prl-102-030502, pachos-njp-11-083010}. Such experiments
are based on postselection of measurements~\citep[as in teleportation
  by][]{bouwmeester-nat-390-575}, which should be contrasted to our
deterministic implementation of open system dynamics to prepare and manipulate
the corresponding quantum state~\citep[as in deterministic teleportation by
][]{riebe-nature-429-734,barrett-nature-429-737}.
  
\subsection{Pumping into ``excited'' states}

Starting from an initially fully mixed state of four qubits, we also
implemented cooling into a different GHZ-type state,
$(\ket{0010}-\ket{1101})/\sqrt{2}$, by a sequence of four dissipative steps: 1)
pumping into the +1 eigenspace of $Z_1Z_2$, 2) pumping into the -1 eigenspace
of $Z_2Z_2$, 3) pumping into the -1 eigenspace of $Z_3Z_4$ and 4) pumping into
the -1 eigenspace of $X_1X_2X_3X_4$. In the context of Kitaev's toric code,
this state would correspond to an excited state. However, as above, we point
out that the underlying Hamiltonian was not implemented in the cooling
dynamics.

\begin{figure}[h!]
\iffigs\begin{center}
\includegraphics{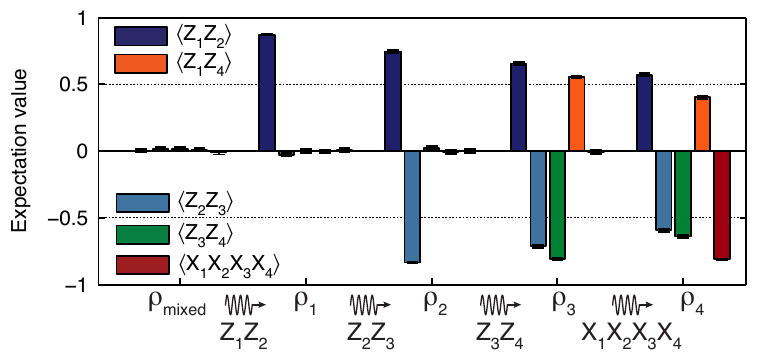}
\end{center}\fi
\caption{\label{fig:heating}\textbf{Cooling into an ``excited'' state.}
  Measured expectation values of two- and four-qubit stabilizer operators for
  pumping into the state $(\ket{0010}-\ket{1101})/\sqrt{2}$, starting from an
  initially four-qubit mixed state.}
\end{figure}

\begin{figure}[h!]
\iffigs \begin{center}
\includegraphics{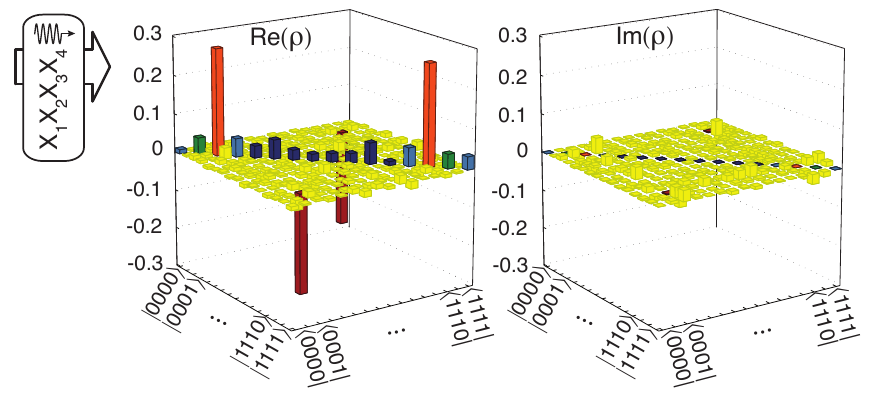}
\end{center}\fi
\caption{\label{fig:pheating}\textbf{Reconstructed density matrix after the
    full cooling cycle for dissipative preparation of the state
    $(\ket{0010}-\ket{1101})/\sqrt{2}$.} This final state has a fidelity of
  60(2)\% with the expected state.  This fidelity was determined from parity
  and coherence measurements and analysed with bayesian inference techniques as
  done in~\cite{monz-superdecoherence}.}
\end{figure}

The measured expectation values of the stabilizers are shown in
Fig.~\ref{fig:heating}. The final density matrix, as determined from quantum
state tomography after the four cooling steps, is shown in
Fig.~\ref{fig:pheating}.  This cooling cycle was implemented with the same
sequences as given for Step 1 to 4 in section \ref{sec:Cooling}, with the only
difference that the sign of the phase shift operations displayed in red was
changed in Steps 2, 3, and 4. This allowed us to invert the pumping direction
from the +1 into -1 eigenspaces of $Z_2Z_2$, $Z_3Z_4$ and $X_1X_2X_3X_4$.

\section{QND measurement of a four-qubit stabilizer}

\subsection{Further details}

\begin{figure}[h!]
\iffigs\begin{center}
\includegraphics{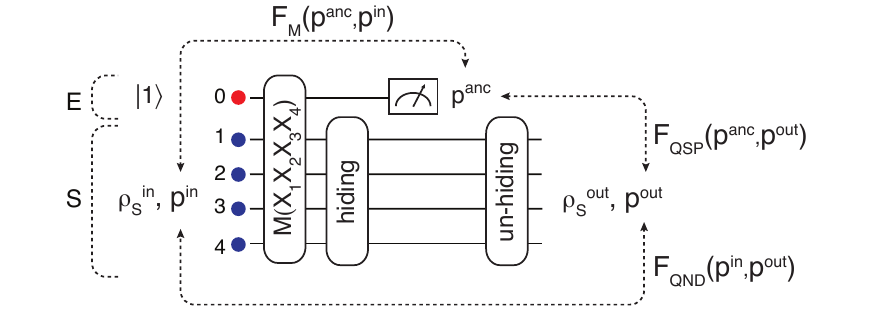}
\end{center}\fi
\caption{\label{fig:FigureQND}\textbf{QND measurement of the four-qubit
    stabilizer operator $X_1X_2X_3X_4$.} After the coherent mapping
  $M(X_1X_2X_3X_4)$, the ancilla qubit is measured. This measurement was
  performed both with and without applying additional pulses to hide the
  populations of the system qubits in electronically uncoupled states for the
  duration of the fluorescence measurement on the ancilla.}
\end{figure}

As shown in Fig.~\ref{fig:FigureQND}, the QND measurement involves a mapping
step where the information about whether the system described by an input
density matrix $\rho^{\mathrm{in}}$ is in the +1 / -1 eigenspace of
$A=X_1X_2X_3X_4$ is coherently mapped onto the internal states $\ket{0}$ and
$\ket{1}$ of the ancilla qubit, which is initially prepared in
$\ket{1}$. Subsequently the ancilla qubit is measured in its computational
basis, leaving the system qubits in a corresponding output state
$\rho^{\mathrm{out}}$.

The coherent mapping $M(X_1X_2X_3X_4)$ was realized by the sequence
\begin{eqnarray*}
U_X(\pi/4)
U_{Z_0}(\pi)\,
U_X(-\pi/4)\\
U_{X^2}(\pi/4)\,U_{X^2}(\pi/4)\,
U_{Z_0}(-\pi/2)\,
U_{X^2}(\pi/4)\,U_{X^2}(\pi/4)\\
U_Y(-\pi/4)\,
U_{Z_0}(\pi)\,
U_Y(\pi/4)\,
\end{eqnarray*}
which implements
\begin{eqnarray}
\label{eq:M}
M(X_1X_2X_3X_4) & = & - \frac{i}{\sqrt{2}}(X_0 + Y_0) \otimes P_+ \nonumber \\
& & \, \, + \frac{1}{\sqrt{2}} (1-i Z_0) \otimes P_-,
\end{eqnarray}
with $P_\pm = \frac 12 (1 \pm X_1X_2X_3X_4)$ the projectors onto the $\pm 1$ eigenspaces of $X_1X_2X_3X_4$. Equation (\ref{eq:M}) shows that for the system qubits being in a state belonging to the +1 eigenspace of the stabilizer operator, the ancilla is flipped from $\ket{1}$ to $\ket{0}$, whereas it remains in its initial state $\ket{1}$ otherwise.

Subsequently, the ancilla as well as the four system qubits were measured. This
was done by measuring the five ions simultaneously. Alternatively, we first hid
the four system qubits in electronic levels decoupled from the laser
excitation, performed the fluorescence measurement of the ancilla qubit, then
recovered the state of the system qubits and tomographically measured the state
of the four system qubits. The second approach, where the state of the system
is not affected by the measurement of the ancilla, is of importance if the
information from the ancilla measurement is to be used for feedback operations
on the state of the system.


\subsection{Quantitative analysis of the performance}

To characterize the performance of a QND measurement for a (multi-)qubit
system, a set of requirements and corresponding fidelity measures have been
discussed in the literature~\citep{ralph-pra-73-012113}.

(1) First of all, the measurement outcomes for the ancilla qubit should agree
with those that one would expect from a direct measurement of the observable $A$
on the input density matrix. This property can be quantified by the {\em
  measurement fidelity},
\begin{equation}
F_\mathrm{M} = \left( \sqrt{p_+^{\mathrm{in}} p_{|0\rangle}^{\mathrm{m}}} +
\sqrt{p_-^{\mathrm{in}} p_{|1\rangle}^{\mathrm{m}}}\right)^2,
\end{equation}
which measures the correlations of the distribution of measurement outcomes
$p^{\mathrm{m}} = \{ p_{|0\rangle}^{\mathrm{m}},
p_{|1\rangle}^{\mathrm{m}} \}$ of the ancilla qubit with the expected
distribution $p^{\mathrm{in}} =\{p_+^{\mathrm{in}}, p_-^{\mathrm{in}}\}$
directly obtained from $\rho^{\mathrm{in}}$, where $p_\pm^{\mathrm{in}} =
\mathrm{Tr}\{ \frac{1}{2}(1\pm A) \rho^{\mathrm{in}}\}$.

(2) The QND character, reflected by the fact that the observable $A$ to be
measured should not be disturbed by the measurement itself, becomes manifest in
ideally identical probability distributions $p^{\mathrm{in}}$ and
$p^{\mathrm{out}}$, which are determined from the input and output density
matrices. These correlations are quantified by the {\em QND fidelity}
\begin{equation}
F_\mathrm{QND} = \left( \sqrt{p_+^{\mathrm{in}} p_+^{\mathrm{out}}} + \sqrt{p_-^{\mathrm{in}} p_-^{\mathrm{out}}}\right)^2, 
\end{equation}
where $p_\pm^{\mathrm{out}} = \mathrm{Tr}\{ \frac{1}{2}(1\pm A)
\rho^{\mathrm{out}}\}$.

(3) Finally, by measuring the ancilla qubit the system qubits should be
projected onto the corresponding eigenspace of the measured observable $A$. Thus
the quality of the QND measurement as a quantum state preparation (QSP) device
is determined by the correlations between the ancilla measurement outcomes and
the corresponding system output density matrices. It can be described by the
{\em QSP fidelity}
\begin{equation}
F_\mathrm{QSP} = p_+^{\mathrm{m}} p_{|0\rangle,+}^{\mathrm{out}} + p_-^{\mathrm{m}} p_{|1\rangle,-}^{\mathrm{out}},
\end{equation}
where $p_{|0/1\rangle,\pm}^{\mathrm{out}}$ denotes the conditional probability
of finding the system qubits in the +1 (-1) eigenspace of $A$, provided the
ancilla qubit has been previously measured in $\ket{0}$ ($\ket{1}$).

The probability distributions for the system input and output states, the
ancilla measurement outcome distributions, and the resulting fidelity values
are summarized in Tables I to IV.  The input states had a
fidelity~\citep{jozsa-jmo-41-2315} with the ideal states
$(|0000\rangle +|1111\rangle)/\sqrt{2}$, $(|0000\rangle-|1111\rangle)/\sqrt{2}$ and $(|0011\rangle-|1100\rangle)/\sqrt{2}$ of
${75.3(9),77.3(8),93.2(4)}$\%.

We observe that we obtain higher values for the measurement and QND fidelities
than for the QSP fidelities. The latter is relevant in the context of quantum
error correction or closed-loop simulation protocols or more generally whenever
the information from the ancilla measurement is used for further processing of
the system output state.

With the additional hiding and unhiding pulses before and after the measurement
of the ancilla we observe a loss of fidelity of a few percent in the QSP
fidelities.

\clearpage

\begin{widetext}

\begin{table}
\caption{QND probability distributions. Obtained from measurements
  \textbf{with} hiding of the system ions during the measurement of the ancilla.}
\begin{tabular}{cccccccccc}                                                                                                                                             
input state& eigenspace &$p^m_{in}$ & $p^m_{out}$ & $p^{in}$ & $p^{in}_{m=0}$ & $p^{in}_{m=1}$ &$p^{out}$ & $p^{out}_{m=0}$ & $p^{out}_{m=1}$ \\       \hline\hline                  
$|0000\rangle+|1111\rangle$
 & $+1$ & 0.959(1) & 0.847(3) & 0.817(9) & 0.822(9) & 0.618(34) & 0.689(12) & 0.736(12) & 0.359(34)\\
 & $-1$ & 0.041(1) & 0.153(3) & 0.183(9) & 0.178(9) & 0.382(34) & 0.311(12) & 0.264(12) & 0.641(34)\\\hline
$|0000\rangle-|1111\rangle$
 & $+1$ & 0.955(1) & 0.169(3) & 0.191(10) & 0.187(9) & 0.328(36) & 0.310(11) & 0.640(26) & 0.242(12)\\
 & $-1$ & 0.045(1) & 0.831(3) & 0.809(10) & 0.813(9) & 0.672(36) & 0.690(11) & 0.360(26) & 0.758(12)\\\hline
$|0011\rangle-|1100\rangle$
 & $+1$ & 0.978(1) & 0.103(2) & 0.041(4) & 0.035(4) & 0.412(47) & 0.137(9) & 0.476(36) & 0.097(7)\\
 & $-1$ & 0.022(1) & 0.897(2) & 0.959(4) & 0.965(4) & 0.588(47) & 0.863(9) & 0.524(36) & 0.903(7)\\\hline
\end{tabular}
\end{table}

\begin{table}
\caption{QND probability distributions. Obtained from measurements
  \textbf{without} hiding of the system ions during the measurement of the ancilla.}
\begin{tabular}{cccccc}
input state& eigenspace & $p^m_{out}$ & $p^{out}$ & $p^{out}_{m=0}$ & $p^{out}_{m=1}$ \\\hline\hline
$|0000\rangle+|1111\rangle$
 & $+1$ & 0.850(3) & 0.713(11) & 0.789(11) & 0.336(30)\\
 & $-1$ & 0.150(3) & 0.287(11) & 0.211(11) & 0.664(30)\\\hline
$|0000\rangle-|1111\rangle$ 
 & $+1$ & 0.188(3) & 0.265(12) & 0.504(28) & 0.220(11)\\
 & $-1$ & 0.812(3) & 0.735(12) & 0.496(28) & 0.780(11)\\\hline
$|0011\rangle-|1100\rangle$
 & $+1$ & 0.099(2) & 0.073(7) & 0.416(35) & 0.038(5)\\
 & $-1$ & 0.901(2) & 0.927(7) & 0.584(35) & 0.962(5)\\\hline
\end{tabular}
\end{table}

\begin{table}
\caption{QND figures of merit. Determined from measurements \textbf{with}
  hiding of the system ions during the measurement of the ancilla.  Since the
  state $|0011\rangle-|1100\rangle$ is particularly robust against decoherence,
  the fidelity $F_\text{QSP}$ is higher, as shown for 8 ions in
  ~\cite{monz-superdecoherence}.}
\begin{tabular}{cccccccccc}
input state& eigenspace &$p^\text{in}$ & $p^\text{out}$ & $p^\text{m}$ & $p^\text{out}_{\text{QND}=+}$
  &$p^\text{out}_{\text{QND}=-}$& $F_\text{M}(p^\text{in},p^\text{m})$ & $F_\text{QND}(p^\text{in},p^\text{out})$ & $F_\text{QSP}(p^\text{m},p^\text{out}_\text{QND})$\\\hline\hline
$|0000\rangle+|1111\rangle$
&$+1$& 0.82(1) & 0.69(1) & 0.85 & 0.74(1) &         & 0.998(1) & 0.978(5) & 0.72(1) \\
&$-1$& 0.18(1) & 0.31(1) & 0.15 &         & 0.64(3) \\\hline
$|0000\rangle-|1111\rangle$
&$+1$& 0.19(1) & 0.31(1) & 0.17 & 0.64(3) &         & 0.999(1) & 0.980(5) & 0.74(1) \\
&$-1$& 0.81(1) & 0.69(1) & 0.83 &         & 0.76(1) \\\hline
$|0011\rangle-|1100\rangle$
&$+1$& 0.041(4)& 0.14(1) & 0.10 & 0.48(4) &         & 0.985(3) & 0.969(6) & 0.86(1) \\
&$-1$& 0.959(4)& 0.86(1) & 0.90 &         & 0.90(1) \\\hline
$|1111\rangle$
&$+1$& 0.5  & 0.47(1) & 0.50049 & 0.70(1) &         & 1        & 0.9992(6)& 0.73(1) \\
&$-1$& 0.5  & 0.53(1) & 0.49951 &         & 0.76(1) \\\hline
\end{tabular}
\end{table}

\begin{table}
\caption{QND figures of merit. Determined from measurements \textbf{without}
  hiding of the system ions during the measurement of the ancilla. Since the
  state $|0011\rangle-|1100\rangle$ is particularly robust against decoherence,
  the fidelity $F_\text{QSP}$ is higher, as shown for 8 ions in
  ~\cite{monz-superdecoherence}.}
\begin{tabular}{ccccccccc}
input state& eigenspace & $p^\text{out}$ & $p^\text{m}$ & $p^\text{out}_{\text{QND}=+}$
  &$p^\text{out}_{\text{QND}=-}$& $F_\text{M}(p^\text{in},p^\text{m})$ & $F_\text{QND}(p^\text{in},p^\text{out})$ & $F_\text{QSP}(p^\text{m},p^\text{out}_\text{QND})$\\\hline\hline
$|0000\rangle+|1111\rangle$
&$+1$& 0.71(1) & 0.85 & 0.79(1) &         & 0.998(1) & 0.984(4) & 0.77(1) \\
&$-1$& 0.29(1) & 0.15 &         & 0.66(3) \\\hline
$|0000\rangle-|1111\rangle$
&$+1$& 0.26(1) & 0.19 & 0.50(3) &         & 1.0000(1) & 0.992(3) & 0.73(1) \\
&$-1$& 0.74(1) & 0.81 &         & 0.78(1) \\\hline
$|0011\rangle-|1100\rangle$
&$+1$& 0.07(1) & 0.10 & 0.42(3) &         & 0.986(2) & 0.996(2) & 0.91(1) \\
&$-1$& 0.93(1) & 0.90 &         & 0.96(1) \\\hline
$|1111\rangle$
&$+1$& 0.52(1) & 0.5078 & 0.75(1) &         & 0.99994& 0.9996(5)& 0.74(1) \\
&$-1$& 0.48(1) & 0.4922 &         & 0.73(1) \\\hline
\end{tabular}
\end{table}

\end{widetext}

\begin{figure*}[h!]
\begin{center}
\iffigs \includegraphics{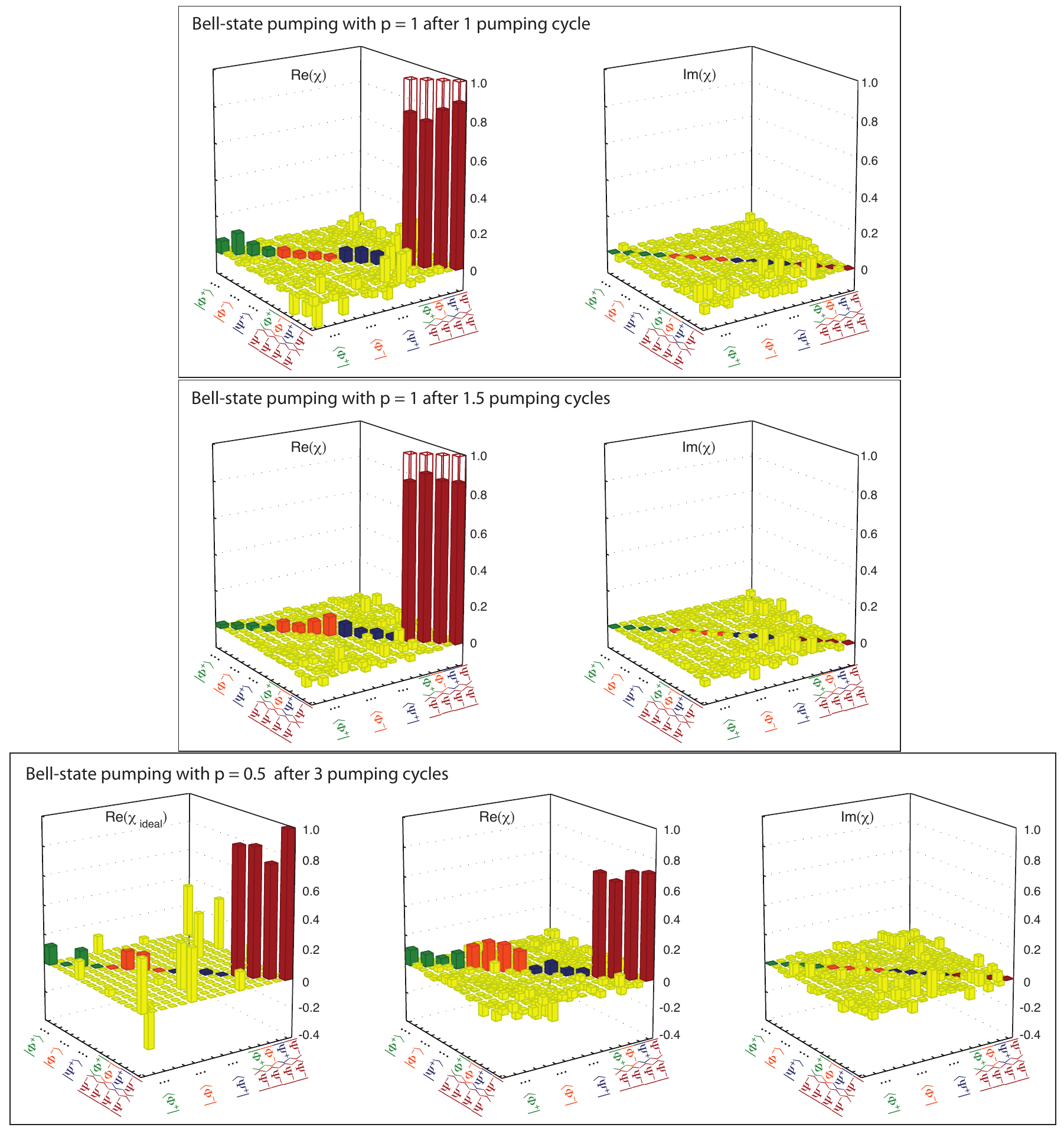}\fi
\end{center}
\caption{\label{fig:bscchis}\textbf{Reconstructed process matrices of
    experimental Bell-state cooling.}  The reconstructed process matrix for
  $p=1$ after 1 (1.5) cycles has a Jamiolkowski process
  fidelity~\cite{gilchrist-pra-71-062310} of 83.4(7)\% (87.0(7)\%) with the
  ideal dissipative process $\rho_\text{S}\mapsto\ket{\Psi^-}\bra{\Psi^-}$
  which maps an arbitrary state of the system into the Bell state
  $\ket{\Psi^-}$.  This ideal process has as non-zero elements only the four
  transparent bars shown.  The reconstructed process matrix for $p=0.5$ after
  3
 cycles has a Jamiolkowski process fidelity of 60(1)\% with the ideal
  process $\chi_\text{ideal}$ shown [$\text{Im}(\chi_\text{ideal})=0$].}

\end{figure*}

\begin{figure*}[h!]
\iffigs\begin{center}
\includegraphics[width=0.7\textwidth]{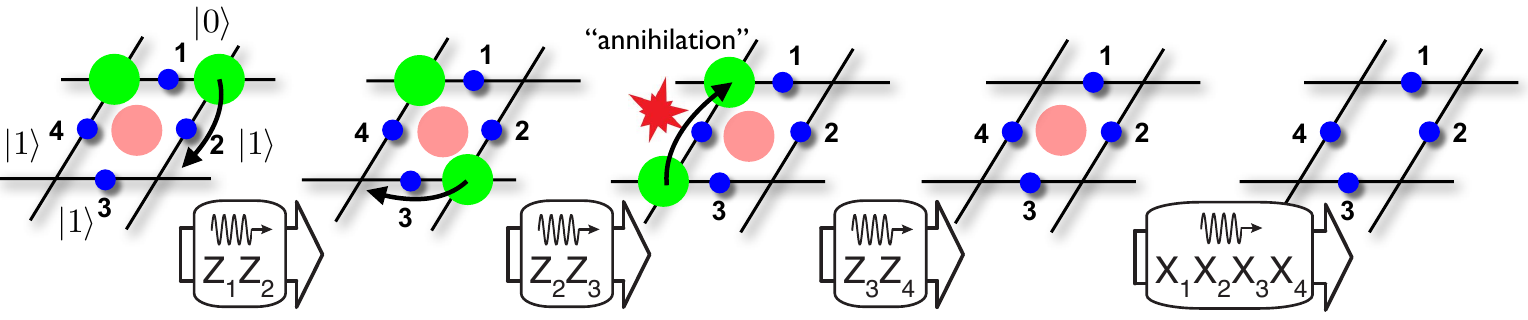}
\end{center}\fi
\caption{\label{fig:AnyonCartoon}\textbf{Pushing ``anyons''.} Cartoon of the
  dissipative dynamics. The pumping dynamics can be visualized by dissipative
  pushing of excitations (green and red dots) between adjacent corners of the
  plaquette.}
\end{figure*}

\bibliography{cooling}
